\documentclass[runningheads]{llncs}

\usepackage{eccv}

\usepackage{eccvabbrv}

\usepackage{graphicx}
\usepackage{booktabs}
\usepackage{multirow}
\usepackage{graphicx}

\usepackage[accsupp]{axessibility}  

\usepackage{hyperref}

\usepackage{orcidlink}

\begin{document}

\title{Phys3DGS: Physically-based 3D Gaussian Splatting for Inverse Rendering} 


\author{Euntae Choi\inst{1} \and
Sungjoo Yoo\inst{1}}

\authorrunning{F.~Author et al.}

\institute{Department of Computer Science and Engineering, Seoul National University, South Korea}

\maketitle

\begin{abstract}
  We propose two novel ideas (adoption of deferred rendering and mesh-based representation) to improve the quality of 3D Gaussian splatting (3DGS) based inverse rendering. We first report a problem incurred by hidden Gaussians, where Gaussians beneath the surface adversely affect the pixel color in the volume rendering adopted by the existing methods. In order to resolve the problem, we propose applying deferred rendering and report new problems incurred in a naive application of deferred rendering to the existing 3DGS-based inverse rendering. In an effort to improve the quality of 3DGS-based inverse rendering under deferred rendering, we propose a novel two-step training approach which (1) exploits mesh extraction and utilizes a hybrid mesh-3DGS representation and (2) applies novel regularization methods to better exploit the mesh. Our experiments show that, under relighting, the proposed method offers significantly better rendering quality than the existing 3DGS-based inverse rendering methods. Compared with the SOTA voxel grid-based inverse rendering method, it gives better rendering quality while offering real-time rendering. 
  \keywords{3D Gaussian splatting \and Inverse rendering \and Regularization}
\end{abstract}

\section{Introduction}
\label{sec:intro}

Inverse rendering has been one of the key problems in computer vision and graphics. Recently, its performance has been significantly improved with the synergy with the neural radiance field (NeRF). The voxel-grid-based NeRF method, TensoIR~\cite{tensoir}, facilitates the decomposition of geometry, material, and lighting. However, it suffers from critical issues of slow rendering speed and difficulties in the decomposition of scenes with complex geometry and appearance. 

In our work, we leverage 3D Gaussian splatting (3DGS) which offers high-quality in view synthesis task compared with the previous voxel grid-based NeRF methods. Especially, its differentiable tile rasterizer gives high rendering speed. There have been proposed inverse rendering methods based on 3DGS~\cite{relit3dgs,gs-ir,gir}. These methods, which we call {\it relightable 3DGS} methods, assign learnable parameters of geometry (opacity and normal) and material (BRDF parameters) to each 3D Gaussian and compute the rendering equation on each Gaussian (to be exact, at the intersection of Gaussian and the ray), and obtain the pixel color via volume rendering over Gaussians intersected with the ray. 

The relightable 3DGS methods exhibit superior rendering quality to NeRF-based inverse rendering methods, especially in scenes with complex geometry, e.g., ficus in the TensoIR-Synthetic dataset. 
However, as our experiments demonstrate, they suffer from significant quality degradation under relighting, i.e., novel lighting conditions. Our analysis shows that the quality degradation is mainly due to the rendering method, adopted in the existing relightable 3DGS models, which is similar to the forward rendering in computer graphics~\cite{rtr_book}.

The existing relightable 3DGS methods first obtain the color of each Gaussian (at the ray intersection) while utilizing the surface normal and BRDF parameters of the Gaussian and calculating the rendering equation under the given lighting condition. Then, they perform volume rendering by a weighted summation of colors on the intersections of Gaussian and ray with the associated opacity of each Gaussian. In such a method, the Gaussians near, but beneath, the surface can influence the ray color. We call those Gaussians hidden Gaussians. 

In principle, the color of hidden Gaussian need not be included in the computation of surface color. However, due to the nature of weighted summation along the ray, the color of hidden Gaussian can affect the ray color, especially when the opacity is not yet fully learned in the early stage of training. More importantly, the color of the hidden Gaussian is obtained with different visibility from that of the surface point due to the location difference between the hidden Gaussian and the real surface point. Thus, its color can be significantly different from the surface color, especially in case of high specularity, which aggravates the problem. See Section \ref{sec:problem} for more details.

We propose a novel physically-based relightable 3DGS (Phys3DGS) method. In order to address the problem of hidden Gaussian, it leverages on {\it deferred rendering}~\cite{rtr_book} where, for each pixel, we calculate the estimated surface point (i.e., the estimated depth from the camera center to the surface along the ray of the pixel) and the associated normal and BRDF parameters at the estimated surface point. Then, we compute the rendering equation with the visibility of the estimated surface point. 

According to our experiments, a naive application of deferred rendering to relightable 3DGS suffers from training instability and difficulties in learning BRDF parameters due to the poor separation of base color and light. We identified these problems result from poor geometry learning and, thus, propose a new mesh-3DGS hybrid scene representation. The mesh helps learn the geometry but also incurs new problems inherent in mesh reconstruction such as sudden change in neighbor vertex locations and abnormally large Gaussian scale in the hybrid representation. To address these problems, we present novel regularization methods which contribute to the quality of learned geometry.

Our experiments demonstrate our proposed method can offer superior rendering quality under new lighting conditions to the existing relightable 3DGS methods and faster and better rendering than the conventional voxel-grid-based inverse rendering method.

\section{Related Work}

\subsection{Relightable NeRF}
PhySG is based on an SDF network and makes all the training procedures differentiable with the help of spherical Gaussian approximation~\cite{physg}. InvRender extends PhySG and improves BRDF decomposition by modeling indirect illumination with an MLP~\cite{invrender}. NeILF improves the modeling of specularity and 2nd-bounce illumination by learning incident lights~\cite{neilf}. TensoIR learns BRDF and surface normal by utilizing the voxel grid structure of TensoRF~\cite{tensorf} and modeling visibility and indirect illumination with the query to its appearance model~\cite{tensoir}.

\subsection{Relightable 3DGS}
GS-IR reduces inference cost by baking occlusion and indirect illumination with cubemap convolution~\cite{gs-ir}.
Relit3DGS builds a BVH with 3D Gaussians to enable point-based ray tracing and also bakes visibility in spherical harmonics for real-time novel view synthesis~\cite{relit3dgs}. GIR models the visibility of Gaussian with a simple test of normal and view vector, and works robustly under the datasets of high specularity by training a CNN for each of diffuse and specular lights~\cite{gir}.

\subsection{Mesh-based Rendering}
NVDiffrec combines deep marching tetrahedra model and neural texture and demonstrates the feasibility of mesh-based end-to-end inverse rendering~\cite{nvdiffrec}.
SuGaR presents a hybrid mesh-Gaussian representation formed by binding Gaussians to each triangle of mesh which is extracted from an initial 3DGS model under SDF-based regularization~\cite{sugar}.
Our work was motivated by SuGar in that, in our method, the mesh is extracted from the initial 3DGS model and each triangle of extracted mesh is assigned a 3D Gaussian. Our key difference is to show that a simple adoption of mesh on the relightable 3DGS model does not work for inverse rendering and offers a novel deferred learning-based two-step training approach which enables more robust and higher quality inverse rendering with the help of our proposed regularization of Gaussian opacity and scale as well as per-Gaussian rotation learning.

\section{Deferred 3DGS Rendering}

\subsection{Hidden Gaussian Problem}
\label{sec:problem}

\begin{figure}[t]
\includegraphics[width=0.7\textwidth]{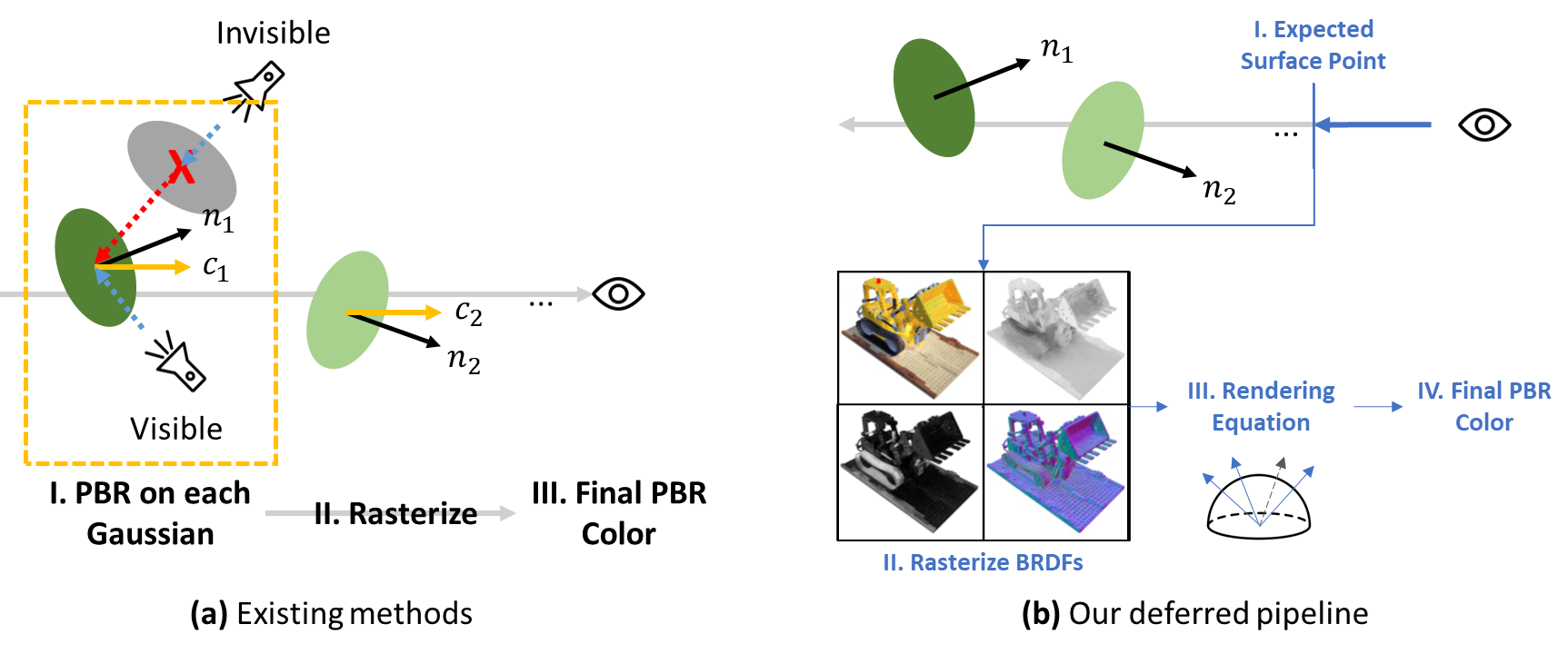}
\centering
\caption{(a) Rendering pipeline of the existing relightable 3DGS models, (b) Our proposed deferred rendering.}
\label{fig:pipeline}
\end{figure}

A 3D Gaussian consists of its center coordinate $\mu$, opacity $\alpha$, scale $s$, 3D covariance $\Sigma$, spherical harmonics (SH) parameters for view-dependent color, normal vector $n$, BRDF (albedo $a$, roughness $r$ and metalness $m$ in our model), and visibility parameters. 
As Figure \ref{fig:pipeline} (a) shows, 
in the existing relightable 3DGS methods, rendering an image follows the flow of physically-based rendering (PBR) for computing per-Gaussian color $c$ (PBR on each Gaussian in the figure), rasterization and final color computation via volume rendering~\cite{relit3dgs,gs-ir}.  
In per-Gaussian color computation, we solve the rendering equation in Eqn. \ref{eqn:renderingeqn} at the ray intersection point of each Gaussian.
The parameters of 3D Gaussian are learned from the photometric loss \cite{3DGS} under normal regularization~\cite{relit3dgs,gs-ir}.

\begin{equation} \label{eqn:renderingeqn}
L(\bm{x}, \bm{\omega}_o) \approx \frac{1}{N} \sum_{i=1}^{N} \frac{L_i(\bm{x}, \bm{\omega}_i)V(\bm{x}, \bm{\omega}_i)f_r(\bm{x}, \bm{\omega}_i, \bm{\omega}_o)(\bm{\omega}_i\cdot\textbf{n})}{p(\bm{\omega}_i)}
\end{equation}

Note that the visibility $V$ in Eqn. \ref{eqn:renderingeqn}, which is learned together with shape and material, is obtained for each Gaussian (center). Thus, as mentioned in Section \ref{sec:intro}, we can have a problem of hidden Gaussian, i.e., a contribution of the color of hidden Gaussians to the pixel color. 

\begin{figure}[h]
\includegraphics[width=1.0\textwidth]{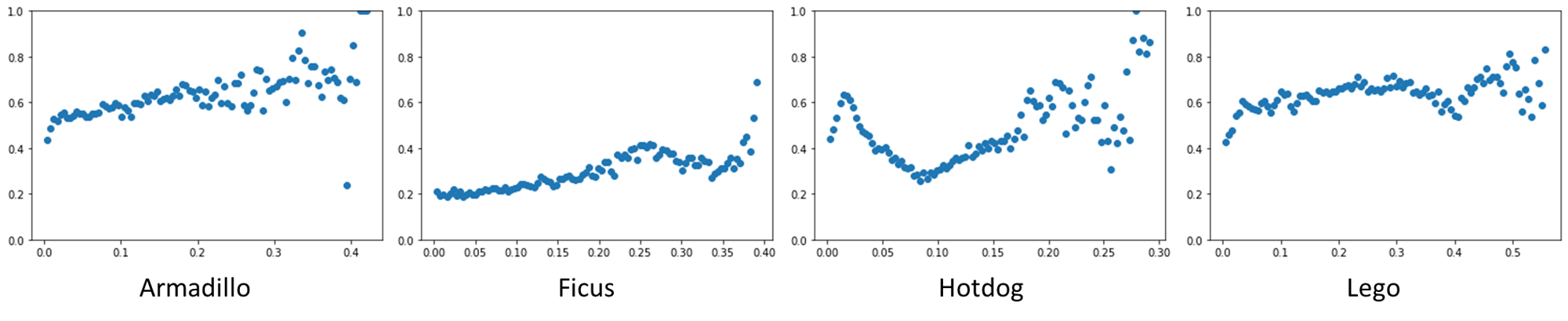}
\centering
\caption{Opacity statistics in the TensoIR-Synthetic scenes.}
\label{fig:opacity}
\end{figure}

Figure \ref{fig:opacity} shows Gaussians tend to have high opacity beneath the surface. We measure the statistics by first obtaining a convex hull from a trained 3DGS model and calculating the distance from the surface of convex hull to the Gaussians inside of the convex hull. The figure shows the average opacity on the Y axis in each bin of distance on the X axis. 


\begin{figure}[h]
\includegraphics[width=1.0\textwidth]{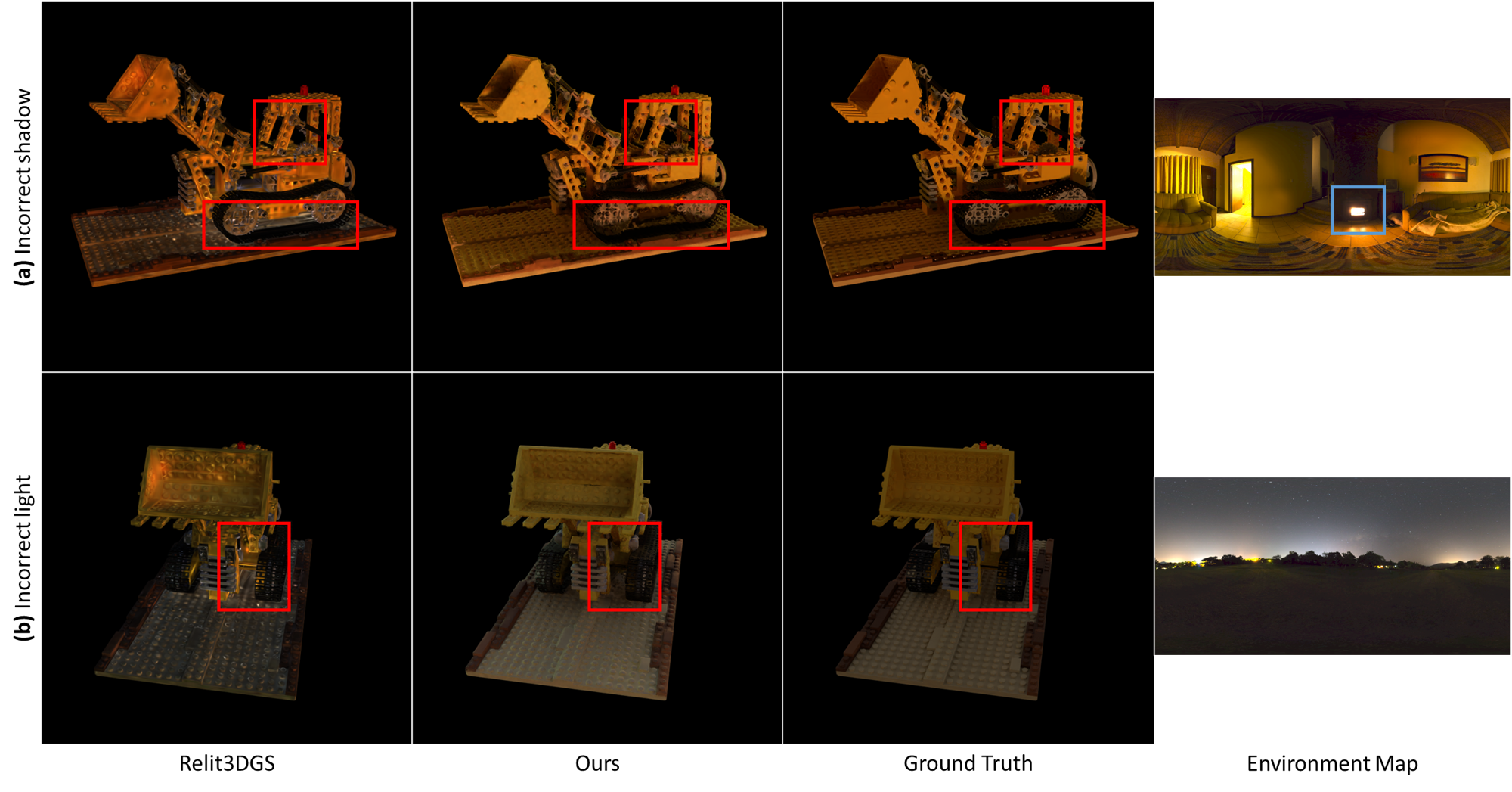}
\centering
\caption{Example scenes with hidden Gaussian problems.}
\label{fig:hiddengaussian}
\end{figure}

Figure~\ref{fig:hiddengaussian} exemplifies the hidden Gaussian problem on the images rendered from a trained 3DGS model. Considering the environment map and its associated rendered image, Figure~\ref{fig:hiddengaussian} (a) shows the rendered image quality can degrade due to incorrectly modeled shadow especially when there is strong incoming light from the left side of lego (a narrow but strong light source highlighted by the blue box is positioned under the lego's base plate). In Figure~\ref{fig:hiddengaussian} (b), even though there is no strong light source from the environment map overall, bright artifacts appear beneath the lego car where a large portion of incoming light should be occluded.


\subsection{Deferred Rendering with Visibility and Light Modeling}

Figure \ref{fig:pipeline} (b) shows our flow of deferred rendering. For each ray (i.e., pixel), we first calculate the estimated depth to localize the estimated surface point by the weighted summation of opacity and distance (to the camera center) of each Gaussian along the ray (Eqn. \ref{eqn:depth}). 

\begin{equation}\label{eqn:depth}
\mathcal{D} = \sum_{i\in N} T_{i} \alpha_i d_i
\end{equation}

In the same manner, we also calculate the estimated normal and BRDF parameters (Eqn. \ref{eqn:brdf} and Rasterize BRDFs in the figure).

\begin{equation}\label{eqn:brdf}
\{\mathcal{A}, \mathcal{M}, \mathcal{R}, \mathcal{N}\} = \sum_{i\in N} T_{i} \alpha_i \{\textbf{a}_i, m_i, r_i, \textbf{n}_i\}
\end{equation}

Then, we solve the rendering equation (Rendering equation in the figure), once for a ray, at the estimated surface point. To compute the incoming light $L(\omega_i)$ at the estimated surface point, we first perform Fibonacci sampling on incoming light directions.

As Eqn. \ref{eqn:4} shows, we model the incoming light with two components, the global and local indirect illumination.

\begin{equation}
\label{eqn:4}
L(\omega_i) = V(\omega_i) \cdot L_{global}(\omega_i) + L_{local}(\omega_i)
\end{equation}

where $V(\omega_i)$ represents the visibility in the direction of $\omega_i$.
For the sampled direction $\omega_i$ of incoming light, we obtain the global illumination $L_{global}(\omega_i)$  by spherical harmonics (SH) approximation.
We also model indirect illumination with SH parameters on each Gaussian and calculate the indirect illumination on the estimated surface point in the same manner as Eqn. \ref{eqn:brdf}. 


We obtain the visibility in two ways. During training, as Figure 1 (b) shows, we obtain the visibility via ray tracing utilizing the bounding volume hierarchy (BVH) structure. 
As the figure shows, the visibility computation is detached from the back-propagation. 
After training, we bake the visibility on each Gaussian.
In order to reduce the model size, we replace the SH coefficients of indirect illumination on each Gaussian with the SH coefficients of visibility obtained after training. 
Such a replacement does not hurt the rendering quality since the SH parameters of indirect illumination are overfit to the training dataset. Thus, in test time, we do not need them under new lighting conditions. Instead, the baked visibility boosts rendering speed by avoiding costly ray tracing in test time.


\section{Phys3DGS: Physically-based 3D Gaussian Splatting}
Our proposed method applies a two-step training approach. 
We first train an initial 3DGS model without physically-based rendering (PBR) capability and obtain a mesh-3DGS hybrid representation (Section \ref{sec:hybrid}).
Then, we train PBR-related parameters on the hybrid representation (Section \ref{sec:PBR}).

\subsection{Mesh-3DGS Hybrid Representation}
\label{sec:hybrid}
We propose exploiting mesh to help 3DGS models better learn the shape.
Our method first obtains an initial 3DGS model and extracts a mesh from it to obtain the mesh-3DGS hybrid representation.
It consists of two stages: mesh extraction after initial training and building a hybrid mesh-3DGS representation.

\subsubsection{Stage 1: Mesh Extraction after Initial Training}
We first train a vanilla 3DGS model as in \cite{sugar}. 
After opacity culling, we obtain an initial 3DGS model. Then, we apply the SDF-based regularization~\cite{sugar} which flattens 3D Gaussians and pushes them toward the surface.
We found that the SDF-based regularization tends to increase the size of Gaussians located near the surface.
Those Gaussians tend to incur two critical problems as follows.
First, mesh extraction produces very large triangles on those large Gaussians, which degrades the quality of the extracted mesh. Second, those large Gaussians interfere with visibility test thereby hurting visibility and incurring gradient explosion on some pixels. 
In order to address these problems, we propose a novel regularization method called {\it masked opacity regularization} defined as follows:
\begin{equation}
L_o = -(1 - M) log(1-o),
\end{equation}
where $M$ is a binary mask whose value is zero on the background and $o$ is the rasterized opacity map containing the accumulated transmittance for each pixel.

Exploiting the mask given in the training image is necessary for high-quality inverse rendering. Typically, we apply mask regularization to push the opacity of Gaussians inside of the mask towards one while pushing that of Gaussians outside of the mask toward zero. We found such a conventional approach can hurt the training of BRDF parameters and adversely affect visibility test. Thus, we regularize the opacity to make the opacity outside of the mask become zero while not constraining the opacity of Gaussians inside of the mask. 
Note that the proposed mask opacity regualization is applied throughout the entire training process (in two stages to obtain the hybrid representation in Section \ref{sec:hybrid} and in the PBR training in Section \ref{sec:PBR}).

In order to extract a mesh, we apply the same procedure of \cite{sugar}, i.e., first obtain a level set from the trained 3DGS model and apply Poisson reconstruction to obtain the surface mesh. We also apply mesh cleanup like duplicates removal.

\begin{figure}[t]
\includegraphics[width=0.7\textwidth]{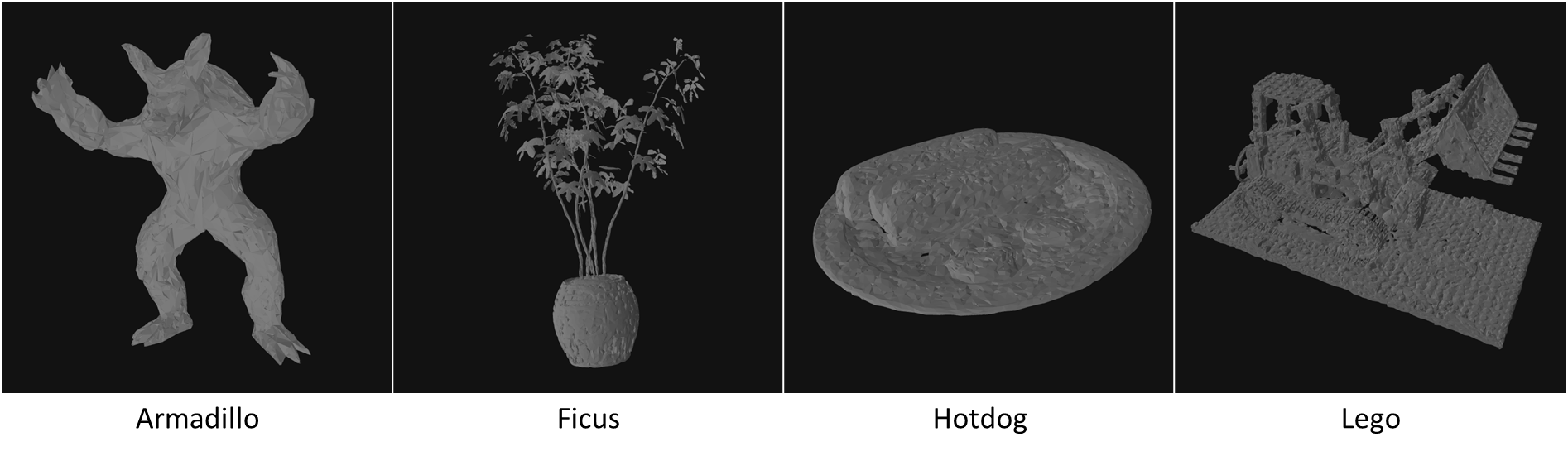}
\centering
\caption{Visualization of extracted surface meshes on TensoIR-synthetic scenes.}
\label{fig:mesh}
\end{figure}

\subsubsection{Stage 2: Building a Mesh-3DGS Hybrid Representation}\label{sec:stage2}
We allocate one flat Gaussian to each triangle of extracted mesh. We call the mesh having one Gaussian on each triangle {\it mesh-3DGS hybrid representation}. After obtaining the hybrid representation, we train the non-PBR parameters of the mesh-3DGS hybrid representation, i.e., mesh geometry (=Gaussian location and shape), opacity, radiance SH coefficients, and scale. 

\begin{figure}[t]
\includegraphics[width=0.7\textwidth]{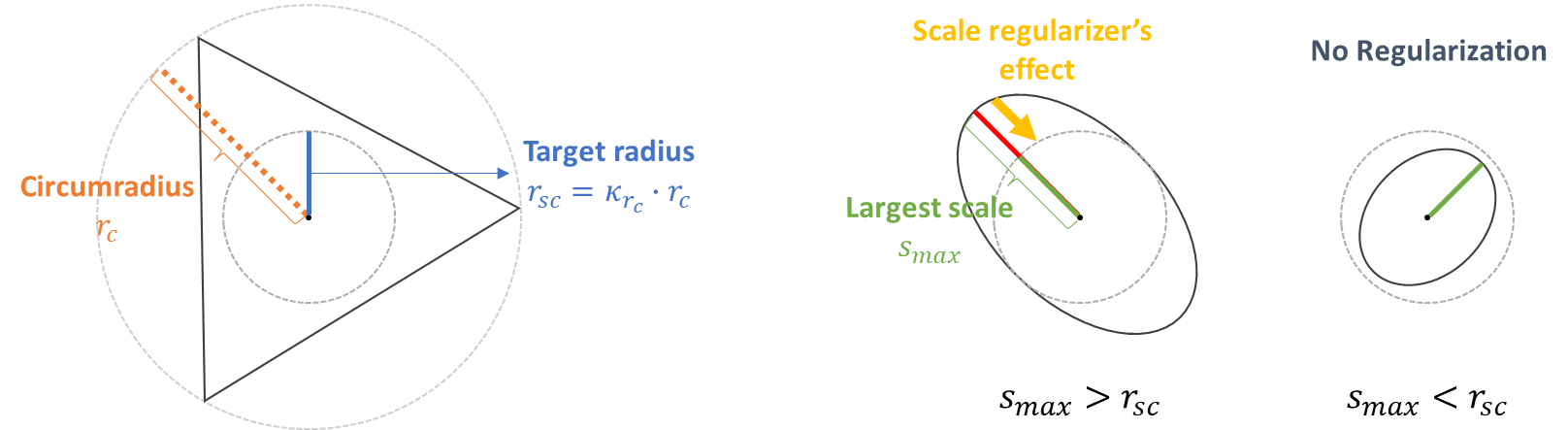}
\centering
\caption{Illustration of our proposed scale regularization loss. For a hypothetical mesh triangle, if the largest scale of a bound Gaussian exceeds the scaled circumradius (or target radius, denoted as $r_{sc}$), our regularizer enforces the Gaussian to reduce its scale value.}
\label{fig:r_sc}
\end{figure}

In the training of the hybrid representation, we identified two critical cases, (A) almost transparent large Gaussians and (B) limited shape learning requiring larger meshes.
In case (A), the large Gaussians can result from (1) the originally large triangles or (2) the training of triangle-mapped Gaussian's scale.
The mask opacity regularization, explained before, can help reduce the occurrence of large triangles of (1), but does not help solve (2).
In order to address the problem of large Gaussians, we propose {\it Gaussian scale regularization}. Specifically, we force the largest scale of Gaussian (with scales in three directions) to be within a threshold of the triangle's scaled circumradius with the following loss function:

\begin{equation}\label{eqn:r_sc}
L_{sc} = \sum_{i\in{N_G}}\textrm{max}(\textrm{max}(s^i) - \kappa_{r_c} r^{i}_{sc}, 0),
\end{equation}
where $N_G$ is the total number of Gaussians, $r^{i}_{sc}$ is the circumradius for $i$-th Gaussian and $\kappa_{r_c}$ is a constant scale set to 0.2.






Our hybrid model has the potential to learn high-quality normal thanks to the adopted mesh. However, utilizing the normal of mesh face as that of its associated Gaussian, like \cite{sugar}, tends to end up requiring a large number of triangles, especially in case of a non-flat surface of the target scene, which corresponds to case (B) mentioned above. In order to further improve the quality of normal and avoid case (B), we propose a {\it quaternion-based normal enhancement} which additionally trains a rotation parameter, a quaternion, on each Gaussian on the mesh surface. Our experiments show the utility of the additional learnable parameter which stabilizes vertex locations and improves the quality of the learned normal vector.

\begin{table}[h]
\caption{Novel view synthesis and relighting results on TensoIR-Synthetic benchmark. All values are in PSNR.}
\label{tab:synthetic}
\resizebox{\textwidth}{!}{%
\begin{tabular}{l|ccccc|ccccc}
\hline
                         & \multicolumn{5}{c|}{Novel View Synthesis}                                                                                                                                               & \multicolumn{5}{c}{Relighting}                                                                                                                                                          \\ \cline{2-11} 
\multirow{-2}{*}{Method} & Armadillo                      & Ficus                          & Hotdog                         & \multicolumn{1}{c|}{Lego}                           & Avg.                           & Armadillo                      & Ficus                          & Hotdog                         & \multicolumn{1}{c|}{Lego}                           & Avg.                           \\ \hline
NeRFactor                & 26.479                         & 21.664                         & 26.479                         & \multicolumn{1}{c|}{26.076}                         & 25.175                         & 26.887                         & 20.684                         & 22.713                         & \multicolumn{1}{c|}{23.246}                         & 23.383                         \\
InvRender                & 31.116                         & 22.131                         & 31.832                         & \multicolumn{1}{c|}{24.391}                         & 27.368                         & 27.814                         & 20.330                         & \cellcolor[HTML]{FFFFB2}27.630 & \multicolumn{1}{c|}{20.117}                         & 23.973                         \\
NVDiffrec                & 33.664                         & 22.131                         & 34.903                         & \multicolumn{1}{c|}{30.056}                         & 30.189                         & 23.099                         & 17.260                         & 19.075                         & \multicolumn{1}{c|}{20.088}                         & 19.881                         \\
TensoIR                  & 39.050                         & 29.780                         & \cellcolor[HTML]{FFD9B2}36.820 & \multicolumn{1}{c|}{34.700}                         & 35.088                         & \cellcolor[HTML]{FFB2B2}34.504 & 24.296                         & \cellcolor[HTML]{FFB2B2}27.927 & \multicolumn{1}{c|}{\cellcolor[HTML]{FFB2B2}28.581} & \cellcolor[HTML]{FFD9B2}28.827 \\ \hline
Relit3DGS                & \cellcolor[HTML]{FFD9B2}41.825 & \cellcolor[HTML]{FFB2B2}37.629 & 36.377                         & \multicolumn{1}{c|}{\cellcolor[HTML]{FFB2B2}36.655} & \cellcolor[HTML]{FFD9B2}38.121 & \cellcolor[HTML]{FFFFB2}31.283 & \cellcolor[HTML]{FFFFB2}29.124 & 22.365                         & \multicolumn{1}{c|}{24.158}                         & 26.732                         \\
GIR                      & \cellcolor[HTML]{FFB2B2}43.379 & \cellcolor[HTML]{FFD9B2}37.352 & \cellcolor[HTML]{FFB2B2}37.916 & \multicolumn{1}{c|}{\cellcolor[HTML]{FFD9B2}36.269} & \cellcolor[HTML]{FFB2B2}38.792 & 27.472                         & \cellcolor[HTML]{FFB2B2}31.927 & 23.331                         & \multicolumn{1}{c|}{\cellcolor[HTML]{FFFFB2}24.261} & \cellcolor[HTML]{FFFFB2}26.748 \\
Ours                     & \cellcolor[HTML]{FFFFB2}41.795 & \cellcolor[HTML]{FFFFB2}30.065 & \cellcolor[HTML]{FFFFB2}36.704 & \multicolumn{1}{c|}{\cellcolor[HTML]{FFFFB2}35.984} & \cellcolor[HTML]{FFFFB2}36.137 & \cellcolor[HTML]{FFD9B2}33.632 & \cellcolor[HTML]{FFD9B2}30.824 & \cellcolor[HTML]{FFD9B2}27.688 & \multicolumn{1}{c|}{\cellcolor[HTML]{FFD9B2}26.417} & \cellcolor[HTML]{FFB2B2}29.640 \\ \hline
\end{tabular}%
}
\end{table}

\subsection{PBR Training}
\label{sec:PBR}
We now also train the PBR-related parameters (albedo or base color, metalness, roughness, and indirect illumination SH coefficients) following our proposed deferred rendering pipeline. The loss functions utilized in Stage 2 are all adopted and the bilateral smoothness loss widely adopted in optimization-based inverse rendering models is applied to all BRDF parameters. 

However, due to the imperfect quality of the surface mesh and large gradients generated by PBR training that is done in HDR space, we still observe degenerate cases where some vertices undergo dramatic position shifts, impairing the quality of visibility modeling, even with the strong regularizers mentioned in Section \ref{sec:stage2}. Therefore, we additionally introduce surface regularization loss as follows:
\begin{equation}\label{eqn:sr}
L_{sr} = ||\mu - \mu_{init}||^2_2,
\end{equation}
where $\mu_{init}$ is the initial position of Gaussian bound to the extracted mesh stored separately. This function enforces Gaussians to stay near their original mesh triangle, which in turn through back-propagation, prevents mesh vertices from getting large updates.

To summarize, our model is trained to minimize the following multi-task loss:
\begin{equation}
L_{total} = L_{L1} + L_{PBR} + \lambda_{smooth} L_{smooth} + \lambda_{o} L_{o} + \lambda_{sc} L_{sc} + \lambda_{sr} L_{sr},
\end{equation}
where $L_{L1}$ and $L_{PBR}$ represent volume rendering loss and photometric loss of PBR, respectively, and $\lambda$s are hyperparameters to set the relative importance of each regularize.


\begin{figure}[h]
\includegraphics[width=1.0\textwidth]{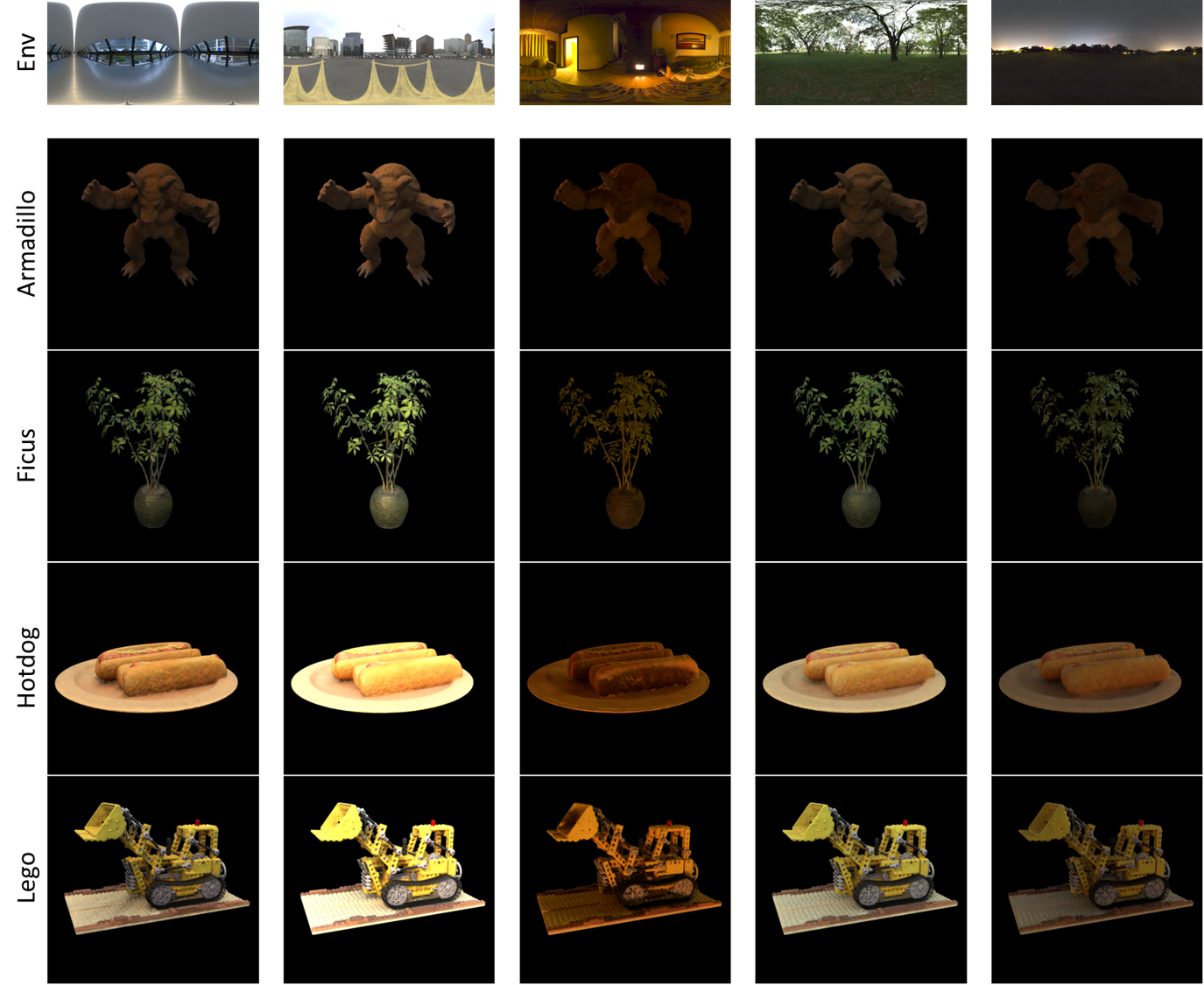}
\centering
\caption{Relighting results on TensoIR-Synthetic benchmark. The environment maps in the first row are mapped to LDR.}
\label{fig:synthetic}
\end{figure}

\section{Experiments}
\subsection{Training \& Testing Details}
Stage 1 (Section \ref{sec:hybrid}) takes 40k steps (30k for vanilla 3DGS and 10k for the SDF regularization) and finishes within an hour on average, including the mesh extraction. When extracting a mesh, we match the number of triangles by adjusting mesh decimation parameters to the number of Gaussians in the corresponding Relit3DGS checkpoint to make a fair comparison to the best we can. Stage 2 runs for 10k steps and takes up to 10 minutes. The PBR training takes 10k steps and the duration is about 30 minutes.

We mostly follow NeILF~\cite{neilf} for evaluating the BRDF function and employ Fibonacci sampling for random incoming light sampling during training. When relighting, we adopt NVDiffrecMC's~\cite{nvdiffrecmc} multiple importance sampling~\cite{veach} by casting an equal amount of light importance rays and GGX importance rays. Also, a non-parametric bilateral denoiser~\cite{svgf} post-processes the PBR result to cope with low samples per point, which is essential to real-time rendering.
Base color and metallic are always clamped to have values between 0 and 1. Roughness is scaled and shifted to lie between 0.09 and 1 to prevent pixels from having exceedingly strong specular reflectance.

All experiments are conducted with a single NVIDIA RTX 4090 GPU running on NGC Docker image \footnote{\url{https://catalog.ngc.nvidia.com/orgs/nvidia/containers/pytorch}}. For more detailed information on our experimental setting, refer to supplementary material.

\begin{figure}[h]
\includegraphics[width=1.0\textwidth]{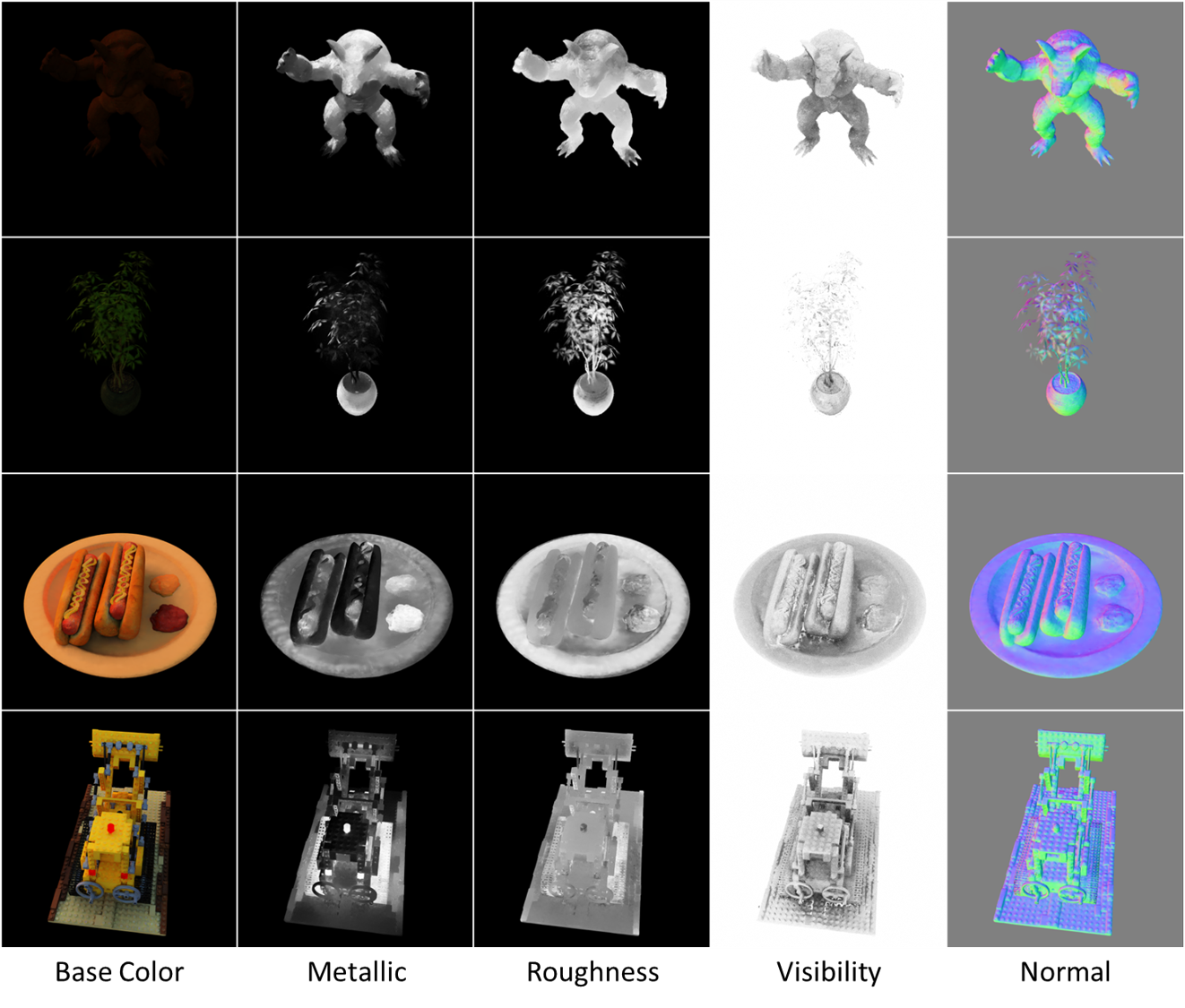}
\centering
\caption{Qualitative evaluation on scene decomposition. }
\label{fig:synthetic_brdf}
\end{figure}

\subsection{Synthetic Scenes}
Table \ref{tab:synthetic} shows a comparison on novel view synthesis and relighting. Novel view synthesis utilizes the same lighting condition as that of training images while relighting utilizes new lighting conditions (5 HDR environment maps in the experiments) to render images.

In novel view synthesis, both of the existing relightable 3DGS methods, Relit3DGS~\cite{relit3dgs} and GIR~\cite{gir} give the best results. 
Ours gives competitive results against the previous voxel grid-based methods like TensoIR. We analyze the quality difference between ours and Relit3DGS and GIR comes from (1) our usage of mesh which constrains the location and orientation of Gaussians and (2) overfit behavior of the existing methods to the training dataset as will be shown in the relighting case below.

In the table, ours offers the best rendering quality in relighting. In particular, ours outperforms, by a large margin, the two relightable 3DGS methods, Relit3DGS and GIR. Considering that relighting needs to take unseen lighting conditions, this result demonstrates the two existing methods tend to overfit to the lighting condition of the training dataset.
Compared with TensorIR which does not run in real-time due to a large number of samples, i.e., 1,024 samples per pixel (spp) for rendering, ours gives better rendering quality with a much fewer spp of 64 and non-parametric denoiser thereby enabling real-time rendering.

A detailed comparison of relighting results helps understand the characteristics of each method. In Ficus, all three relightable 3DGS methods including ours outperform the best voxel grid-based method by a large margin (24.296 vs 29.124 to 31.927), which demonstrates the advantage of 3DGS methods on complex geometry.
In Armadillo and Hotdong with simple overall geometry but non-flat surface, ours offers comparable quality to the voxel grid-based method while the other relightable 3DGS methods give poor rendering quality.
In Lego (having a flat surface), the voxel grid-based method outperforms all three relightable 3DGS methods. The high quality of the voxel grid-based method is mostly due to the large number of samples per pixel at the penalty of slow rendering speed. On the contrary, ours offers a better trade-off between rendering quality (best in relighting) and speed (real-time).

Figure \ref{fig:synthetic} exemplifies relighting results with various environment maps on the TensoIR-Synthetic benchmark. Figure \ref{fig:synthetic_brdf} gives a qualitative evaluation on scene decomposition. In the figure, the visibility represents the hit test result averaged over sampled light directions, which accounts for ambient occlusion.

\begin{table*}[t]
\centering
\caption{Novel view synthesis results on real-world DTU dataset. }
\label{tab:real}
\resizebox{\linewidth}{!}{
\begin{tabular}{lcccccccccccccccc}
\hline
DTU Scan  & 24                                                & 37                                                & 40                                                & 55                                                & 63                                                & 65                                                & 69                                                & 83                                                & 97                                                & 105                                               & 106                                               & 110                                               & 114                                               & 118                                               & 122                                               & Avg.                                              \\ \hline
NerFactor & 23.24                                             & 21.91                                             & 23.33                                             & 26.86                                             & 22.70                                             & 24.71                                             & 27.59                                             & 22.56                                             & 20.45                                             & 25.08                                             & 26.30                                             & 25.14                                             & 21.35                                             & 26.44                                             & 26.53                                             & 24.28                                             \\
PhySG     & 17.38                                             & 15.11                                             & 20.65                                             & 18.71                                             & 18.89                                             & 18.47                                             & 18.08                                             & 21.98                                             & 17.31                                             & 20.67                                             & 18.75                                             & 17.55                                             & 21.20                                             & 18.78                                             & 23.16                                             & 19.11                                             \\
Neual-PIL & 20.67                                             & 19.51                                             & 19.12                                             & 21.01                                             & 23.70                                             & 18.94                                             & 17.05                                             & 20.54                                             & 19.88                                             & 19.67                                             & 18.20                                             & 17.75                                             & 21.38                                             & 21.69                                             & —                                                 & 19.94                                             \\
Nvdiffrec & 18.86                                             & 20.90                                             & 20.82                                             & 19.46                                             & 25.70                                             & 22.82                                             & 20.34                                             & \cellcolor[HTML]{FFFFB2}25.17                     & 20.16                                             & 24.21                                             & 20.01                                             & 21.56                                             & 20.85                                             & 22.81                                             & 25.02                                             & 21.91                                             \\
NeILF++   & \cellcolor[HTML]{FFD9B2}27.31                     & \cellcolor[HTML]{FFB2B2}26.21                     & \cellcolor[HTML]{FFB2B2}28.19                     & \cellcolor[HTML]{FFFFB2}30.07                     & \cellcolor[HTML]{FFFFB2}27.47                     & \cellcolor[HTML]{FFFFB2}26.79                     & \cellcolor[HTML]{FFB2B2}30.92                     & 24.63                                             & \cellcolor[HTML]{FFFFB2}24.56                     & \cellcolor[HTML]{FFD9B2}29.25                     & \cellcolor[HTML]{FFD9B2}31.58                     & \cellcolor[HTML]{FFD9B2}30.69                     & \cellcolor[HTML]{FFFFB2}26.93                     & \cellcolor[HTML]{FFFFB2}31.33                     & \cellcolor[HTML]{FFFFB2}33.19                     & \cellcolor[HTML]{FFFFB2}28.61                     \\ \hline
Relit3DGS & \cellcolor[HTML]{FFB2B2}27.87                     & \cellcolor[HTML]{FFFFB2}24.98                     & \cellcolor[HTML]{FFD9B2}27.31                     & \cellcolor[HTML]{FFB2B2}32.02                     & \cellcolor[HTML]{FFB2B2}30.89                     & \cellcolor[HTML]{FFB2B2}31.22                     & \cellcolor[HTML]{FFD9B2}29.45                     & \cellcolor[HTML]{FFB2B2}31.88                     & \cellcolor[HTML]{FFB2B2}27.05                     & \cellcolor[HTML]{FFB2B2}29.63                     & \cellcolor[HTML]{FFB2B2}34.36                     & \cellcolor[HTML]{FFB2B2}32.94                     & \cellcolor[HTML]{FFB2B2}30.54                     & \cellcolor[HTML]{FFB2B2}36.76                     & \cellcolor[HTML]{FFB2B2}37.31                     & \cellcolor[HTML]{FFB2B2}30.95                     \\
Ours      & \multicolumn{1}{l}{\cellcolor[HTML]{FFFFB2}25.89} & \multicolumn{1}{l}{\cellcolor[HTML]{FFD9B2}25.42} & \multicolumn{1}{l}{\cellcolor[HTML]{FFFFB2}27.21} & \multicolumn{1}{l}{\cellcolor[HTML]{FFD9B2}30.85} & \multicolumn{1}{l}{\cellcolor[HTML]{FFD9B2}30.52} & \multicolumn{1}{l}{\cellcolor[HTML]{FFD9B2}31.10} & \multicolumn{1}{l}{\cellcolor[HTML]{FFFFB2}29.31} & \multicolumn{1}{l}{\cellcolor[HTML]{FFD9B2}30.58} & \multicolumn{1}{l}{\cellcolor[HTML]{FFD9B2}25.76} & \multicolumn{1}{l}{\cellcolor[HTML]{FFFFB2}28.60} & \multicolumn{1}{l}{\cellcolor[HTML]{FFFFB2}31.50} & \multicolumn{1}{l}{\cellcolor[HTML]{FFFFB2}30.12} & \multicolumn{1}{l}{\cellcolor[HTML]{FFD9B2}29.19} & \multicolumn{1}{l}{\cellcolor[HTML]{FFD9B2}34.94} & \multicolumn{1}{l}{\cellcolor[HTML]{FFD9B2}34.47} & \multicolumn{1}{l}{\cellcolor[HTML]{FFD9B2}29.73} \\ \hline
\end{tabular}%
}
\end{table*}

\subsection{Real Scenes}

Table \ref{tab:real} compares the quality of novel view synthesis (NVS) on real scenes in the DTU dataset. 
Though our model performs slightly worse in NVS compared to Relit3DGS, it outperforms neural network-based models.
Note that it is not possible to evaluate relighting performance since the DTU dataset does not contain any test images captured under different light conditions.

Tables \ref{tab:synthetic} and \ref{tab:real} give similar trends in the comparison of NVS performance between the existing relightable 3DGS methods and ours.
The existing relightable 3DGS methods tend to offer better NVS performance (possibly due to the overfitting issue mentioned in the case of relighting results in Table \ref{tab:synthetic}).
We would expect similar results in the case of relighting experiments with real scenes as we observed in Table 
\ref{tab:synthetic}.



%


\begin{table}[h]
\centering
\caption{Trade-off between rendering speed and average rendering quality in terms of samples per pixel.}
\label{tab:speed}
\resizebox{0.25\textwidth}{!}{%
\begin{tabular}{@{}ccc@{}}
\toprule
SPP & FPS   & Avg. PSNR \\ \midrule
32  & 72.54 & 29.292    \\
48  & 53.19 & 29.303    \\
64  & 42.75 & 29.640    \\ \bottomrule
\end{tabular}%
}
\end{table}

\subsection{Ablation Studies}

\begin{figure}[h]
\includegraphics[width=0.4\textwidth]{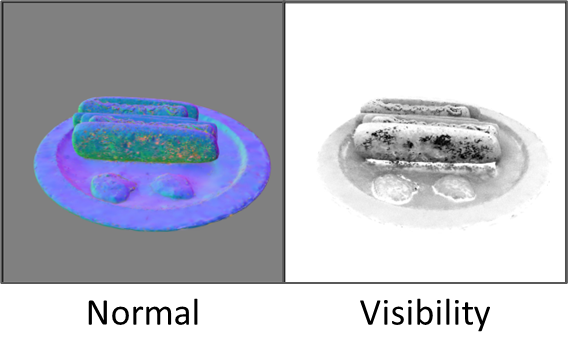}
\centering
\caption{Qualitative result exhibiting ill-reconstructed surface normal and visibility when our proposed normal rotation is not employed.}
\label{fig:normalrotation}
\end{figure}


A comparison of Figures \ref{fig:synthetic_brdf} and  \ref{fig:normalrotation} exemplifies the effect of quaternion-based normal enhancement (Section \ref{sec:hybrid}).
When the additional quaternion to learn normal rotation is excluded, our model easily becomes unstable so that gradient explosion occurs at about 30\% of the total training process. As shown in Figure \ref{fig:normalrotation}, adjusting the vertex positions alone (to learn rotation) is not able to correctly reconstruct the surface geometry. This causes some pixels to be occluded in almost all directions (black artifacts in the visibility map), which makes the outgoing radiance zero in extreme cases.

Table \ref{tab:speed} shows the trade-off between rendering speed and quality. While increasing rendering speed by more than double from 24.75 to 72.54 FPS, it has only a small loss (29.640 to 29.292 dB) of rendering quality, which demonstrates its robustness to fewer numbers of samples per pixel.

Refer to supplementary material for additional ablation results on our proposed regularizers, light modeling, and a naive deferred rendering of Relit3DGS~\cite{relit3dgs}.

%

\section{Limitations and Future Work}
Our method is based on surface mesh and, thus, has a limitation in modeling translucency. Allocating additional non-mesh-bound Gaussians inside of mesh under differentiable path tracing looks promising as a future work for translucency.
The quality of our model strongly depends on that of the extracted mesh. Thus, it is required to evaluate our proposed methods on more complex and real scenes to uncover their limitations, especially, on self-intersection or imperfect surface coverage. In this context, diffusion-based mesh generation with the initially trained 3DGS model as a condition looks promising for better mesh extraction. 


\section{Conclusion}
In this work, we reported the hidden Gaussian problem in the existing relightable 3DGS methods and proposed applying deferred rendering. We also reported the problems of a naive application of deferred rendering.
As the solution, we presented a two-step training approach that utilizes a hybrid mesh-3DGS representation for better shape learning. In order to resolve the problems inherent in the hybrid representation and enable robust training, we proposed masked opacity regularization, Gaussian scale regularization, and quaternion-based normal enhancement.
To cope with the problem of gradient explosion induced by the large value range of PBR training in HDR space, we further propose surface regularization where the initially extracted surface mesh can serve as a strong prior for object geometry.
Our experiments on relighting show that ours outperforms the existing relightable 3DGS methods by a large margin. Compared with the SOTA voxel grid-based inverse rendering method, ours offers real-time rendering as well as better rendering quality in relighting.

\pagebreak

%
%
\bibliographystyle{splncs04}
\bibliography{main}
\end{document}


\title{Supplementary Material for Phys3DGS: Physically-based 3D Gaussian Splatting for Inverse Rendering}

\maketitle

\section{Additional Experiments}

\begin{figure}[]
\includegraphics[width=1.0\textwidth]{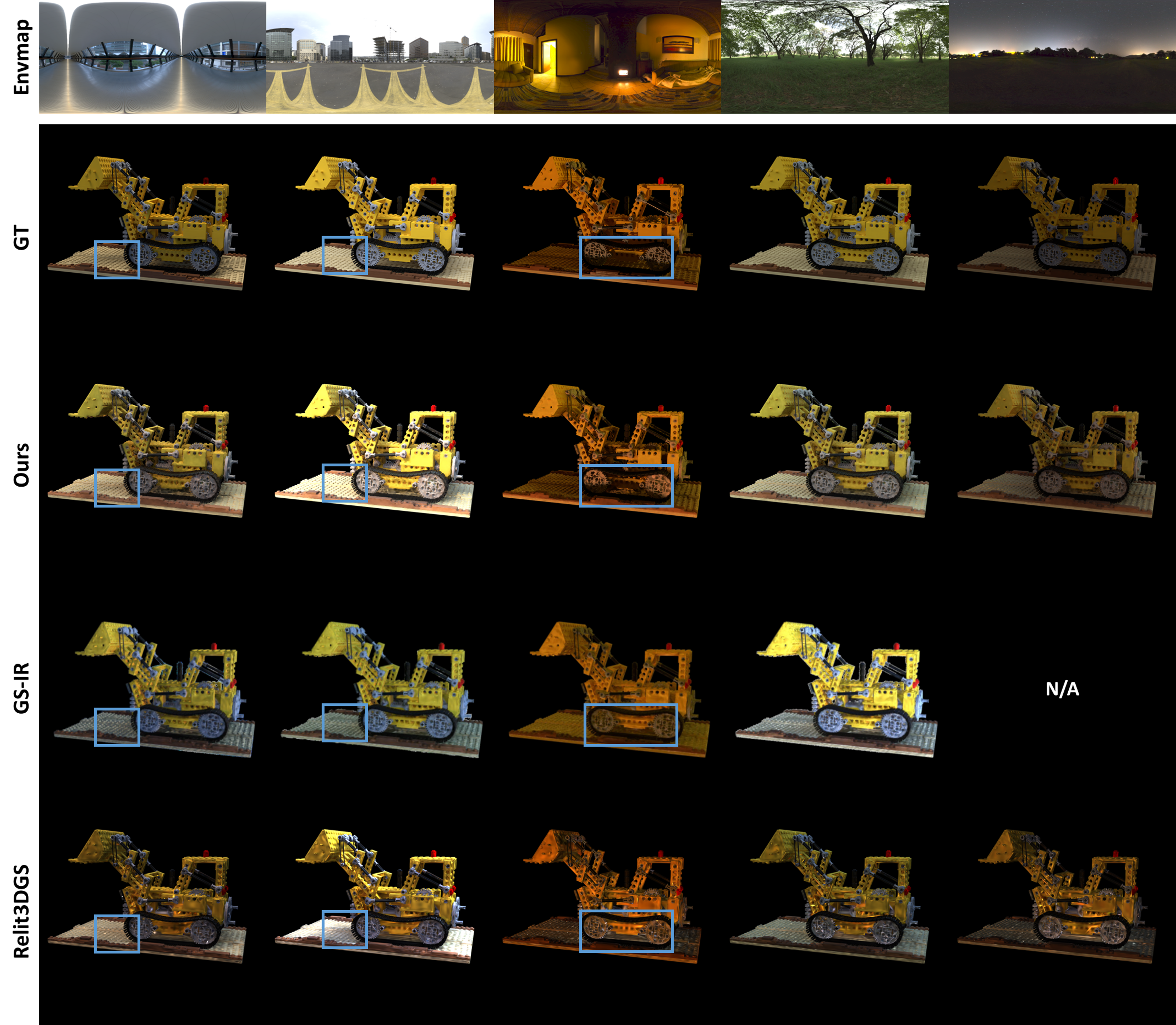}
\centering
\caption{Relighting comparison on TensoIR-Synthetic lego scene.}
\label{fig:legorelit}
\end{figure}

\subsection{More Results on TensoIR-Synthetic Scenes}
\begin{figure}[]
\includegraphics[width=0.8\textwidth]{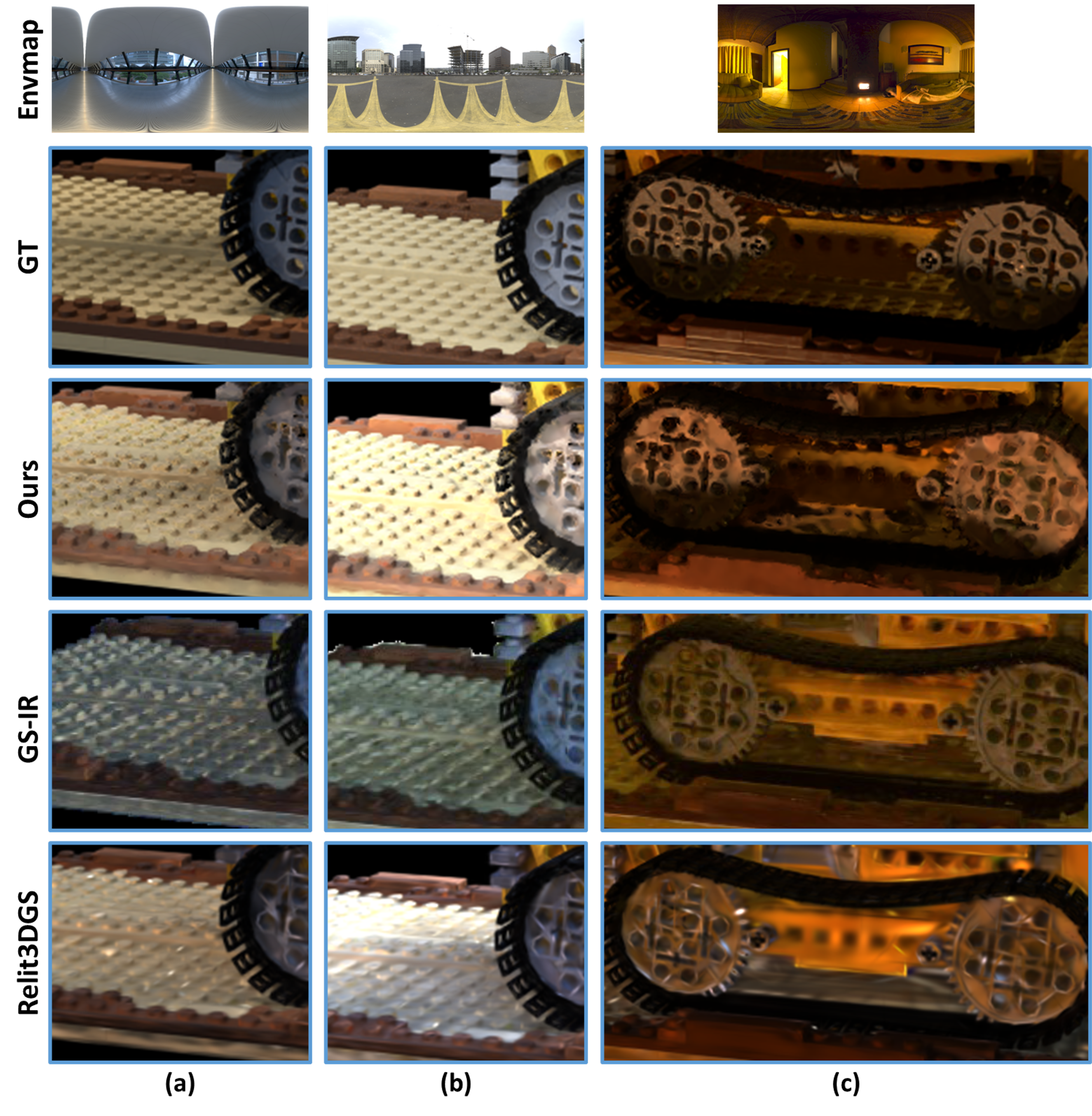}
\centering
\caption{Enlarged details on TensoIR-Synthetic lego scene for clearer comparison.}
\label{fig:legodetail}
\end{figure}

Fig. \ref{fig:legorelit} compares our model to ground truth and other relightable 3DGS models~\cite{gs-ir,relit3dgs}. Relit3DGS suffers from the hidden Gaussian problem, resulting in speckled rendering and incorrect shadow modeling. In Fig. \ref{fig:legodetail}, we show the enlarged images of highlighted boxes in Fig. \ref{fig:legorelit} that clearly demonstrate such issues. In Fig. \ref{fig:legodetail} (a) and (c), only ours renders shadow in the correct area, and in Fig. \ref{fig:legodetail} (c), Relit3DGS gives wrong shadow.

\begin{figure}[]
\includegraphics[width=1.0\textwidth]{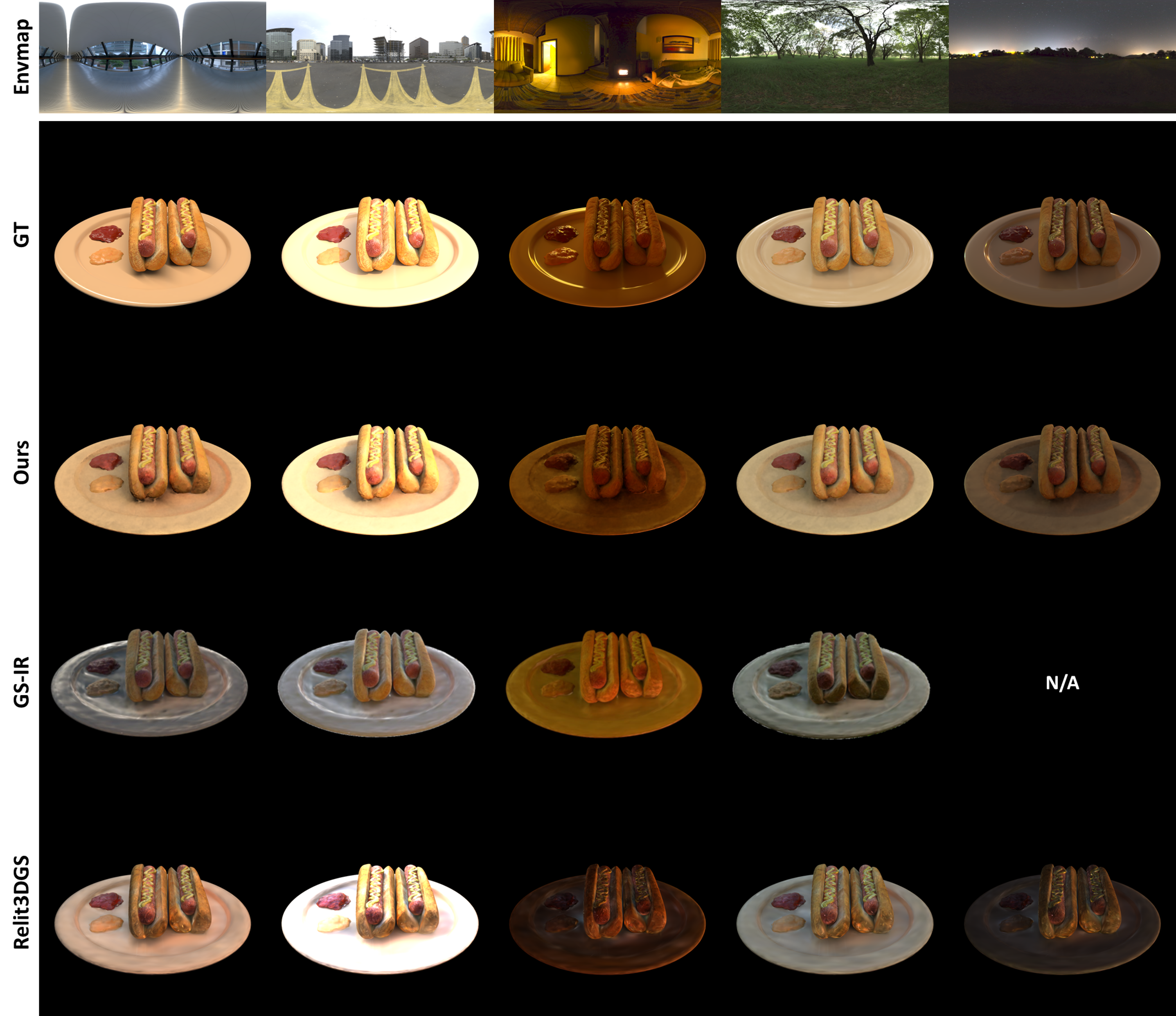}
\centering
\caption{Relighting comparison on TensoIR-Synthetic hotdog scene.}
\label{fig:hotdogrelit}
\end{figure}

For the hotdog scene shown in Fig. \ref{fig:hotdogrelit}, our model generates the closest rendering to ground truth. Specifically, GS-IR fails to model the flat surface of the dish and both existing approaches render incorrect or no shadow. However, ours is also not successful in modeling strong specular light on the edges of the dish. We analyze that, in our rendering pipeline, all pixels in a single image are considered as a minibatch, thus the samples per point (SPP) value is significantly limited to avoid OOM errors and ensure fast real-time rendering. In future work, the pipeline can be reworked such that tiny patches from many images are drawn and the SPP is set to 512 or 1k as in TensoIR~\cite{tensoir}, which helps reduce the variance of sampled light while allowing us to utilize the 3DGS rasterizer without modification.

In Table \ref{tab:syntheticdetail}, we compare ours to GS-IR and other neural inverse rendering methods on NVS and relighting tasks with three metrics (PSNR, SSIM, and LPIPS).

\begin{table}[]
\centering
\caption{Qualitative comparison on TensoIR-Synthetic scenes.}
\label{tab:syntheticdetail}
\resizebox{0.7\textwidth}{!}{%
\begin{tabular}{lc|ccc|ccc}
\hline
                        &                          & \multicolumn{3}{c|}{Novel View Synthesis}                                                      & \multicolumn{3}{c}{Relight}                                                                    \\
\multirow{-2}{*}{Scene} & \multirow{-2}{*}{Method} & PSNR                           & SSIM                          & LPIPS                         & PSNR                           & SSIM                          & LPIPS                         \\ \hline
                        & NeRFactor                & 26.076                         & 0.881                         & 0.151                         & 23.246                         & \cellcolor[HTML]{FFFFB2}0.865 & 0.156                         \\
                        & InvRender                & 24.391                         & 0.883                         & 0.151                         & 20.117                         & 0.832                         & 0.171                         \\
                        & NVDiffrec                & 30.056                         & \cellcolor[HTML]{FFFFB2}0.945 & 0.059                         & 20.088                         & 0.844                         & \cellcolor[HTML]{FFFFB2}0.114 \\
                        & TensoIR                  & \cellcolor[HTML]{FFD9B2}34.700 & \cellcolor[HTML]{FFD9B2}0.968 & \cellcolor[HTML]{FFFFB2}0.037 & \cellcolor[HTML]{FFB2B2}28.581 & \cellcolor[HTML]{FFB2B2}0.944 & \cellcolor[HTML]{FFB2B2}0.081 \\
                        & GS-IR                    & \cellcolor[HTML]{FFFFB2}34.379 & \cellcolor[HTML]{FFD9B2}0.968 & \cellcolor[HTML]{FFD9B2}0.036 & \cellcolor[HTML]{FFFFB2}23.256 & 0.842                         & 0.117                         \\
\multirow{-6}{*}{Lego}  & Ours                     & \cellcolor[HTML]{FFB2B2}35.984 & \cellcolor[HTML]{FFB2B2}0.975 & \cellcolor[HTML]{FFB2B2}0.024 & \cellcolor[HTML]{FFD9B2}26.417 & \cellcolor[HTML]{FFD9B2}0.901 & \cellcolor[HTML]{FFD9B2}0.089 \\ \hline
Hotdog                  & NeRFactor                & 24.498                         & 0.940                         & 0.141                         & 22.713                         & 0.914                         & 0.159                         \\
                        & InvRender                & 31.832                         & 0.952                         & 0.089                         & \cellcolor[HTML]{FFFFB2}27.630 & \cellcolor[HTML]{FFFFB2}0.928 & \cellcolor[HTML]{FFB2B2}0.089 \\
                        & NVDiffrec                & \cellcolor[HTML]{FFFFB2}34.903 & \cellcolor[HTML]{FFFFB2}0.972 & 0.054                         & 19.075                         & 0.885                         & 0.118                         \\
                        & TensoIR                  & \cellcolor[HTML]{FFB2B2}36.820 & \cellcolor[HTML]{FFD9B2}0.976 & \cellcolor[HTML]{FFD9B2}0.045 & \cellcolor[HTML]{FFB2B2}27.927 & \cellcolor[HTML]{FFD9B2}0.933 & \cellcolor[HTML]{FFFFB2}0.115 \\
                        & GS-IR                    & 34.116                         & \cellcolor[HTML]{FFFFB2}0.972 & \cellcolor[HTML]{FFFFB2}0.049 & 21.572                         & 0.888                         & 0.140                         \\
                        & Ours                     & \cellcolor[HTML]{FFD9B2}36.704 & \cellcolor[HTML]{FFB2B2}0.977 & \cellcolor[HTML]{FFB2B2}0.035 & \cellcolor[HTML]{FFD9B2}27.688 & \cellcolor[HTML]{FFB2B2}0.940 & \cellcolor[HTML]{FFD9B2}0.090 \\ \hline
Armadillo               & NeRFactor                & 26.479                         & 0.947                         & 0.095                         & 26.887                         & 0.944                         & 0.102                         \\
                        & InvRender                & 31.116                         & 0.968                         & 0.057                         & \cellcolor[HTML]{FFFFB2}27.814 & \cellcolor[HTML]{FFFFB2}0.949 & \cellcolor[HTML]{FFFFB2}0.069 \\
                        & NVDiffrec                & 33.664                         & \cellcolor[HTML]{FFD9B2}0.983 & \cellcolor[HTML]{FFB2B2}0.031 & 23.099                         & 0.921                         & \cellcolor[HTML]{FFD9B2}0.063 \\
                        & TensoIR                  & \cellcolor[HTML]{FFFFB2}39.050 & \cellcolor[HTML]{FFB2B2}0.986 & \cellcolor[HTML]{FFFFB2}0.039 & \cellcolor[HTML]{FFB2B2}34.504 & \cellcolor[HTML]{FFB2B2}0.975 & \cellcolor[HTML]{FFB2B2}0.045 \\
                        & GS-IR                    & \cellcolor[HTML]{FFD9B2}39.287 & 0.980                         & \cellcolor[HTML]{FFFFB2}0.039 & 27.737                         & 0.918                         & 0.091                         \\
                        & Ours                     & \cellcolor[HTML]{FFB2B2}41.795 & \cellcolor[HTML]{FFFFB2}0.982 & \cellcolor[HTML]{FFD9B2}0.035 & \cellcolor[HTML]{FFD9B2}33.632 & \cellcolor[HTML]{FFD9B2}0.957 & \cellcolor[HTML]{FFD9B2}0.063 \\ \hline
Ficus                   & NeRFactor                & 21.664                         & 0.919                         & 0.095                         & 20.684                         & \cellcolor[HTML]{FFFFB2}0.907 & 0.107                         \\
                        & InvRender                & 22.131                         & 0.934                         & 0.057                         & 20.330                         & 0.895                         & \cellcolor[HTML]{FFFFB2}0.073 \\
                        & NVDiffrec                & 22.131                         & 0.946                         & 0.064                         & 17.260                         & 0.865                         & \cellcolor[HTML]{FFFFB2}0.073 \\
                        & TensoIR                  & \cellcolor[HTML]{FFFFB2}29.780 & \cellcolor[HTML]{FFD9B2}0.973 & \cellcolor[HTML]{FFFFB2}0.041 & \cellcolor[HTML]{FFFFB2}24.296 & \cellcolor[HTML]{FFB2B2}0.947 & \cellcolor[HTML]{FFD9B2}0.068 \\
                        & GS-IR                    & \cellcolor[HTML]{FFB2B2}33.551 & \cellcolor[HTML]{FFB2B2}0.976 & \cellcolor[HTML]{FFB2B2}0.031 & \cellcolor[HTML]{FFD9B2}24.932 & 0.893                         & 0.081                         \\
                        & Ours                     & \cellcolor[HTML]{FFD9B2}30.065 & \cellcolor[HTML]{FFFFB2}0.962 & \cellcolor[HTML]{FFD9B2}0.032 & \cellcolor[HTML]{FFB2B2}30.824 & \cellcolor[HTML]{FFD9B2}0.951 & \cellcolor[HTML]{FFB2B2}0.052 \\ \hline
\end{tabular}%
}
\end{table}

\subsection{Qualitative Results on DTU Results}
\begin{figure}[]
\includegraphics[width=1.0\textwidth]{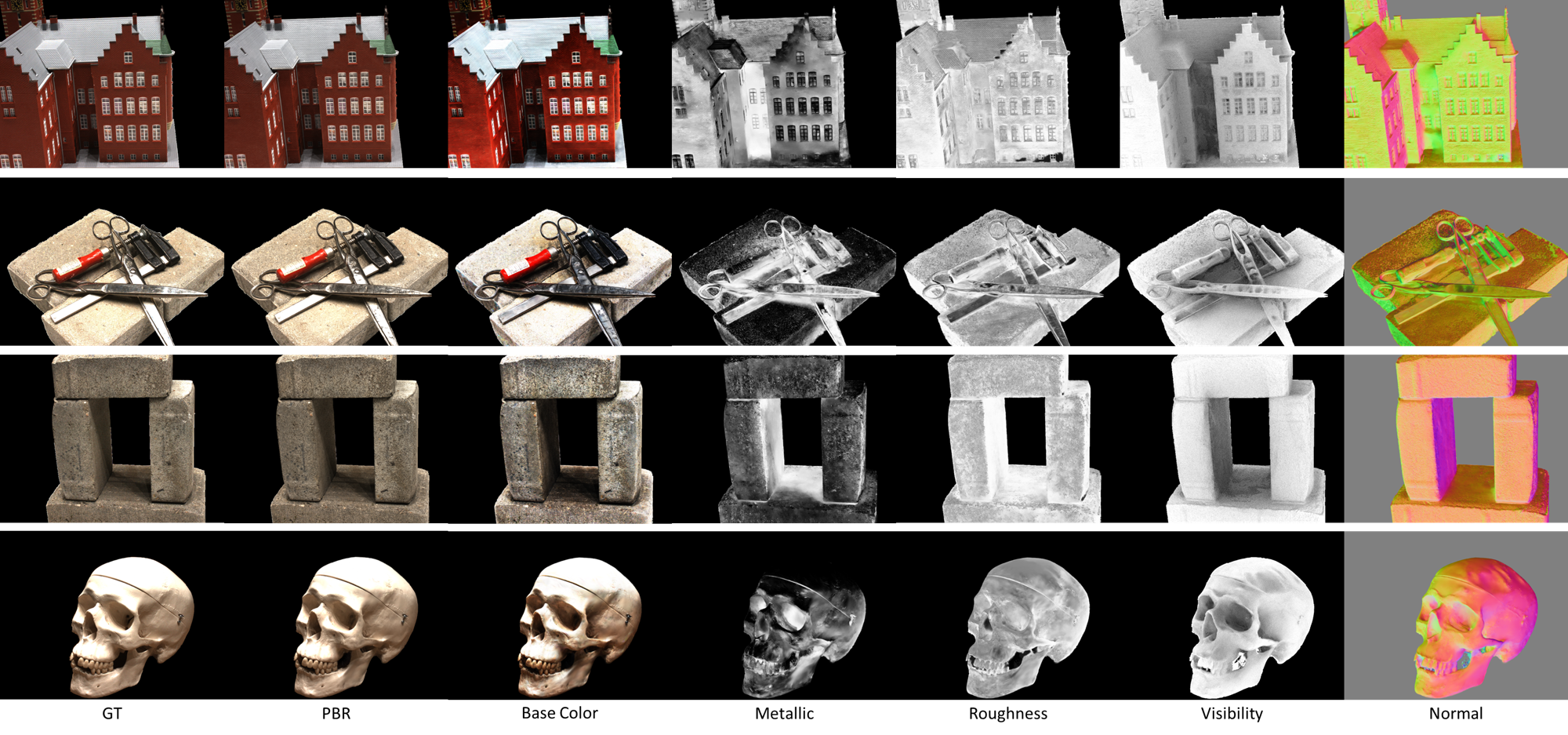}
\centering
\caption{Novel view synthesis and scene decomposition results on 4 scenes in DTU dataset. The scene numbers are 24, 37, 40, and 65.}
\label{fig:dtu}
\end{figure}

We also present qualitative results on DTU scenes in Fig. \ref{fig:dtu}. Note that the samples per point value is set to only 12 to ensure online rendering on an Nvidia RTX 4090 GPU, yet the final PBR rendering results show visually satisfying quality.

\subsection{Results on MipNeRF-360 Unbounded Dataset}

\begin{figure}[]
\includegraphics[width=1.0\textwidth]{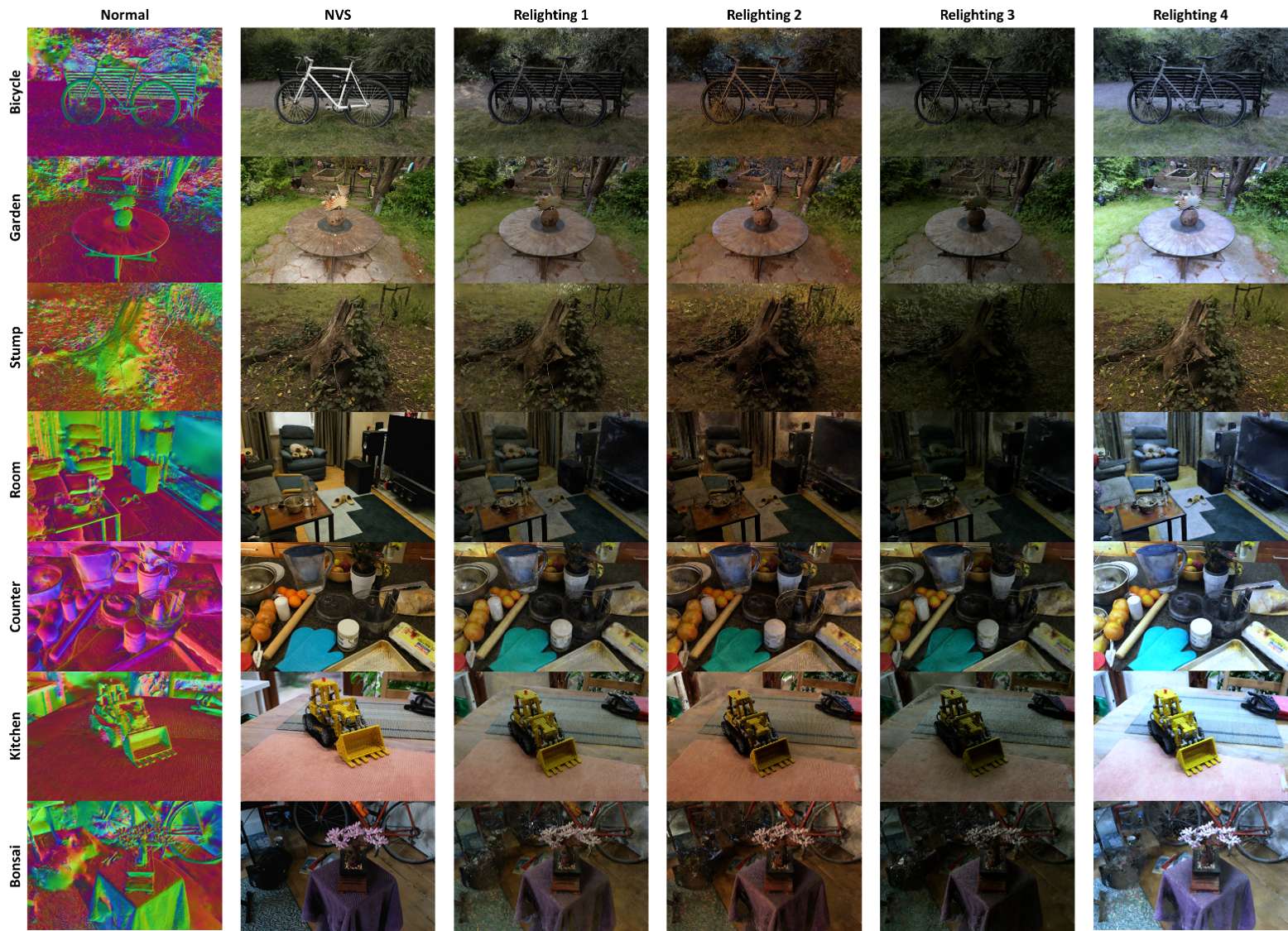}
\centering
\caption{Qualitative results on 7 scenes in MipNeRF-360 dataset.}
\label{fig:mip360}
\end{figure}

\begin{table}[]
\centering
\caption{Novel view synthesis performance evaluated on MipNeRF-360 scenes. All values are in PSNR.}
\label{tab:mip360}
\resizebox{0.8\textwidth}{!}{%
\begin{tabular}{@{}l|ccc|cccc|c@{}}
\toprule
Method    & bicycle                       & garden                        & stump                         & room                          & counter                       & kitchen                       & bonsai                        & Avg.                                                  \\ \midrule
NeRF++~\cite{nerfpp}    & \cellcolor[HTML]{FFFFB2}22.64 & 24.32                         & \cellcolor[HTML]{FFFFB2}24.34 & 28.87                         & \cellcolor[HTML]{FFFFB2}26.38 & 27.80                         & \cellcolor[HTML]{FFFFB2}29.15 & 26.214                                                \\
Plenoxels~\cite{plenoxels} & 21.91                         & 23.49                         & 20.66                         & 27.59                         & 23.62                         & 23.42                         & 24.67                         & 23.625                                                \\
INGP-Base~\cite{instantngp} & 22.19                         & 24.60                         & 23.63                         & \cellcolor[HTML]{FFFFB2}29.27 & \cellcolor[HTML]{FFD9B2}26.44 & \cellcolor[HTML]{FFD9B2}28.55 & \cellcolor[HTML]{FFD9B2}30.34 & 26.430                                                \\
INGP-Big~\cite{instantngp}  & 22.17                         & \cellcolor[HTML]{FFFFB2}25.07 & 23.47                         & \cellcolor[HTML]{FFB2B2}29.69 & \cellcolor[HTML]{FFB2B2}26.69 & \cellcolor[HTML]{FFB2B2}29.48 & \cellcolor[HTML]{FFB2B2}30.69 & \cellcolor[HTML]{FFB2B2}{\color[HTML]{333333} 26.750} \\ \midrule
GS-IR~\cite{gs-ir}     & \cellcolor[HTML]{FFD9B2}23.68 & \cellcolor[HTML]{FFD9B2}25.47 & \cellcolor[HTML]{FFB2B2}25.22 & \cellcolor[HTML]{FFD9B2}29.56 & 26.16                         & \cellcolor[HTML]{FFFFB2}27.90 & 28.65                         & \cellcolor[HTML]{FFFFB2}26.659                        \\
Ours      & \cellcolor[HTML]{FFB2B2}23.98 & \cellcolor[HTML]{FFB2B2}26.46 & \cellcolor[HTML]{FFD9B2}24.83 & 28.94                         & 26.14                         & 27.73                         & 28.63                         & \cellcolor[HTML]{FFD9B2}26.672                        \\ \bottomrule
\end{tabular}%
}
\end{table}

To assess our model's performance on unbounded scenes, we also evaluate it on MipNeRF-360~\cite{mipnerf360} dataset and list the results in Table \ref{tab:mip360}. On average, ours is marginally better than GS-IR~\cite{gs-ir} and outperforms several NeRF-based models. Our model shows impressive results in outdoor scenes (bicycle, garden, and stump) by having an average PSNR gap of 1.52 over INGP-Big~\cite{instantngp} model. However, this is not the case in the indoor scenes. We analyze that it is due to the direct illumination modeling of environment map where all lights are assumed to be coming from an infinitely far hemisphere. Indoor scenes usually generate meshes that surround the scene center. In such cases, almost all incoming light directions for direct illumination are occluded, which significantly lowers the average visibility value, and thus the optimizer is forced to model the entire light relying heavily on the indirect illumination.

During the experiments, we identified that the BVH tracing implementation adopted from Relit3DGS~\cite{relit3dgs} can still produce not-a-number (NaN) results on visibility even with our proposed regularizers so that all parameter values are instantly broken and nothing is rendered. Therefore, we replace the NaNs with zeros while training. 

\subsection{More Ablation Studies}

\subsubsection{Effect of Masked Opacity Regularizer}

\begin{figure}[]
\includegraphics[width=1.0\textwidth]{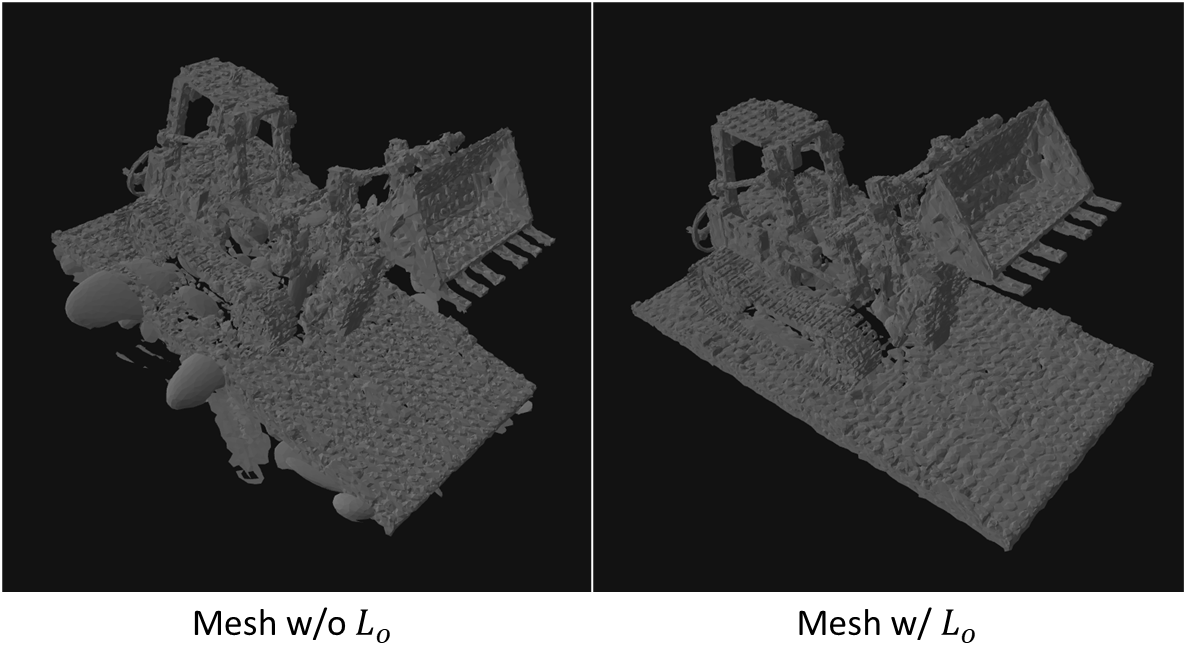}
\centering
\caption{Qualitative comparison on the quality of extracted surface mesh by ablating the masked opacity loss $L_o$.}
\label{fig:l_o}
\end{figure}

When the masked opacity regularization is not applied, the extracted surface mesh is severely degraded as presented in Fig. \ref{fig:l_o}. First, many large ellipsoidal artifacts are generated especially near the object edges since the corresponding Gaussians are free to have influence outside the object region by having small opacity and large scale values. Second, the surface of the mesh becomes rougher overall. 

\subsubsection{Effect of Scale and Surface Regularizer}

\begin{figure}[]
\includegraphics[width=1.0\textwidth]{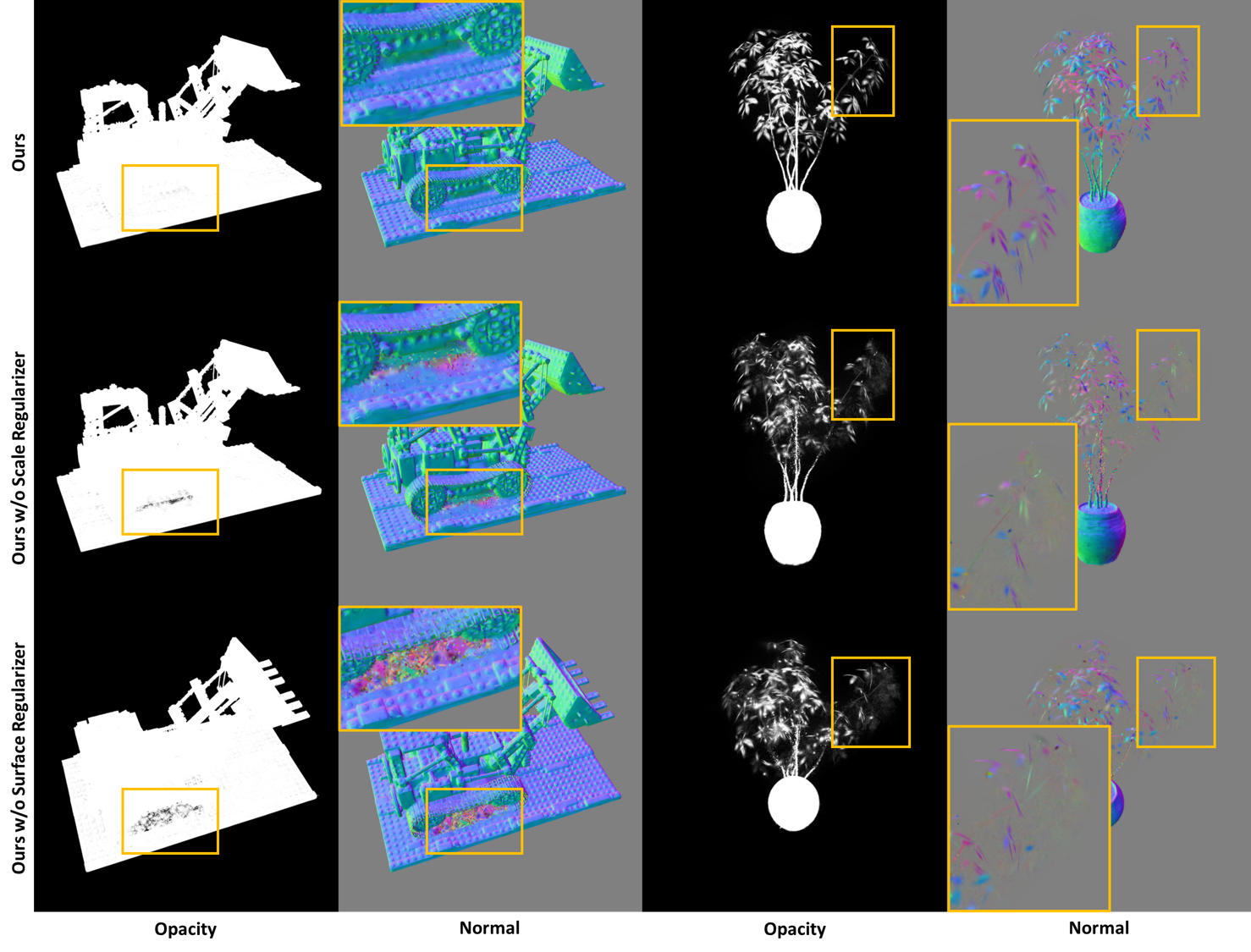}
\centering
\caption{Opacity and normal comparison by ablating our proposed scale regularization loss $L_{sc}$ and surface regularization loss $L_{sr}$.}
\label{fig:l_scsr}
\end{figure}

We ablate our proposed scale and surface regularizer ($L_{sc}$ and $L_{sr}$) to demonstrate their role in training our model. As shown in Fig. \ref{fig:l_scsr}, both loss functions contribute to avoiding errors in opacity and normal modeling. When the scale is not regularized, a lot of Gaussians positioned in regions not covered by a sufficient number of training views tend to pierce outside the surface, and we can observe that they result in chaotic normal modeling in the lego scene. This phenomenon becomes severe when the vertex position is not regularized, leading to indiscriminate mesh structures. In the ficus scene where the mesh contains numerous small parts, poor regularization causes incorrect and disappearing surfaces.

\subsubsection{Effect of Learnable Normal Rotation}

\begin{figure}[]
\includegraphics[width=0.7\textwidth]{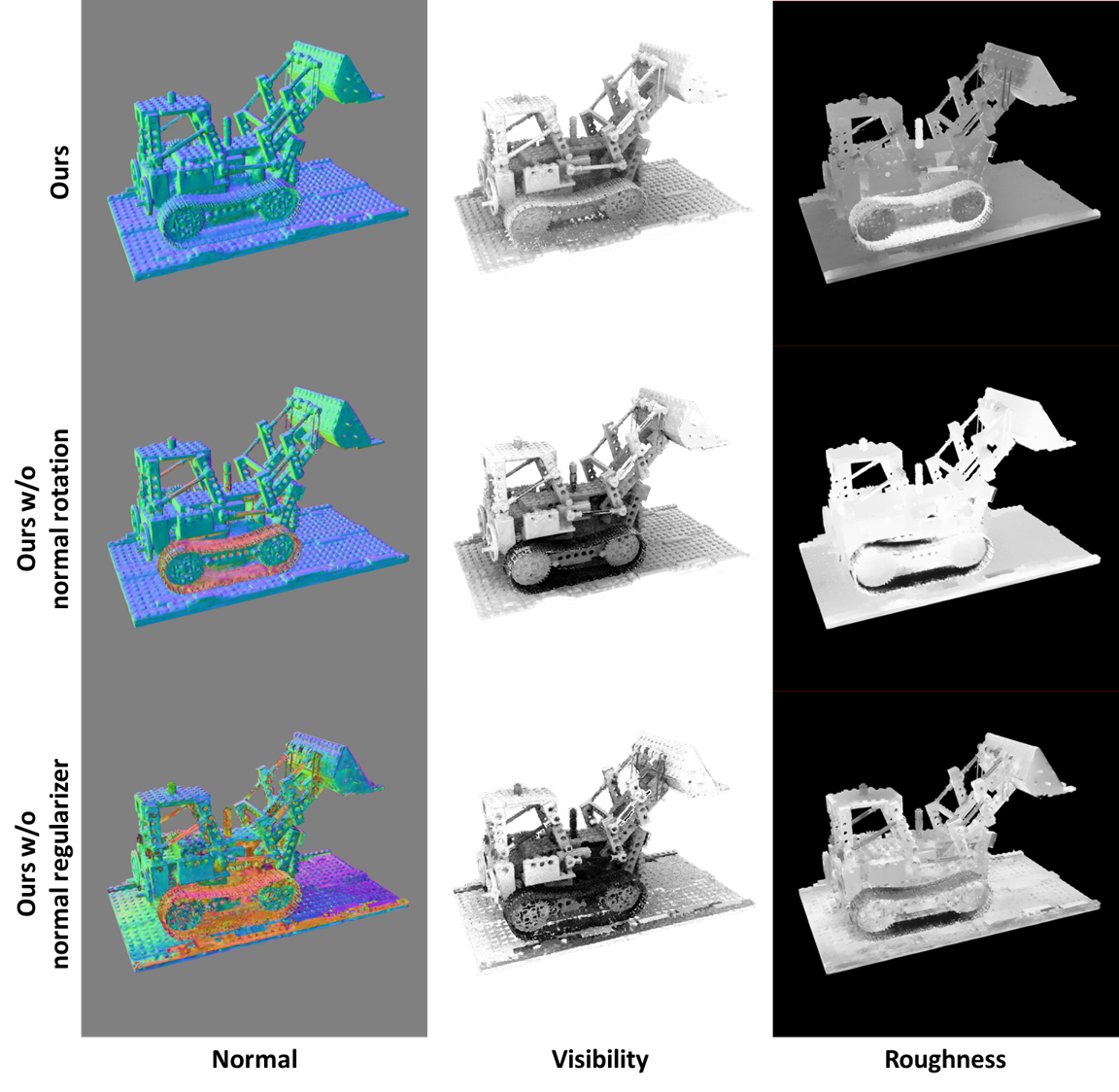}
\centering
\caption{Qualitative comparison to demonstrate the effect of our proposed normal rotation and existing depth gradient-based normal regularization.}
\label{fig:legonormal}
\end{figure}

Fig. \ref{fig:legonormal} shows the effectiveness of our proposed learnable normal rotation modeling. When it is ablated, the rendered normal in some parts of the object is rotated in the wrong direction, and as a side effect, the visibility of such parts becomes nearly zero. Especially, the roughness is also affected by this phenomenon so that visibility-dependent artifacts are generated near the track of the vehicle.


\section{Details on Loss Functions}

\subsection{Normal Regularizer}
We utilize the depth gradient map (a.k.a. pseudo-normal $\bm{n}'$) as a secondary source of geometry prior (other than the surface mesh itself). This is essential for preventing our proposed normal rotation from overfitting to the training light condition (i.e., instead of learning correct geometry and material, it can simply rotate normals to adjust the average amount of incoming lights to minimize the PBR rendering loss as shown in the third row in Fig. \ref{fig:legonormal}).

Since normalized mesh face normals $\bm{n}_{face}$ are $\alpha$-blended by the 3DGS rasterizer, we cannot ensure that the final rendered normals $\bm{n}$ have unit length. Therefore, the rendered normal map is L2-normalized and we maximize cosine similarity to $\bm{n}'$ by minimizing the following loss function:

\begin{equation}
L_{normal}=-\sum_{i\in M}(\frac{\bm{n}_i}{\norm{\bm{n}_i}} \cdot \bm{n}'_i),
\end{equation}
where $M$ is the binary image mask and $i$ is the pixel index on the screen space.

\subsection{Smoothness Regularizer}
Following previous works, we apply image gradient-aware smoothness loss to the BRDF parameters (base color, roughness, and metallic) defined as follows:
\begin{equation}
L_{smooth}(x) = \sum_{i\in M} \sum_{d\in x,y} |\nabla^2_d x(i)| e^{\nabla_d I(i)},
\end{equation}
where $i$ is the pixel index and $I$ is the ground truth RGB image.

\subsection{Scale Regularizer}
The circumradius of a mesh triangle is computed as follows:
\begin{equation}
r_{c} = \frac{\norm{a} \norm{b} \norm{a-b}}{2\norm{a\times b}},
\end{equation}
where $a$ and $b$ are two edge vectors sharing a vertex in the triangle as a starting point. The scale factor $\kappa_{r_c}$ is empirically set to 0.2.

\section{Experimental Settings}

\subsection{Code Release}
A complete codebase will be made public via GitHub upon publication. For now, we attach our core code that makes it possible to run all experiments on TensoIR-Synthetic dataset with the supplementary material. Please unzip \textit{Phys3DGS.zip} included in the archived supplementary file and refer to \textit{instruction.txt}.

\subsection{Computing Environment}
All our experiments (including relightable 3DGS baseline results) are conducted with identical SW and HW environments. We employ an NGC PyTorch Docker container version 23.08 on an Ubuntu 20.04 machine equipped with Nvidia RTX 4090 GPUs. The version of PyTorch and CUDA are 1.10.0 and 11.4, respectively.

\subsection{Mesh Extraction and Number of Gaussians}
We set the depth hyperparameter for Poisson reconstruction to 10. Then, vertices with low-10\% density are pruned by default and quadratic decimation~\cite{quaddecim} is applied to match the desired number of triangles. However, this can be too harsh for simpler scenes such as armadillo and hotdog, possibly making holes in the final mesh. Therefore, the quantile values for these two scenes are set to 0.1\% and 1\%, respectively.

\begin{table}[]
\centering
\caption{Number of Gaussians employed on scenes we experiment on.}
\label{tab:ngaussian}
\resizebox{\textwidth}{!}{%
\begin{tabular}{@{}lcccc|ccccccccccccccc@{}}
\toprule
\multirow{2}{*}{Scene} & \multicolumn{4}{c|}{TensoIR-Synthetic} & \multicolumn{15}{c}{DTU}                                                                             \\ \cmidrule(l){2-20} 
                       & Armadillo   & Ficus  & Hotdog  & Lego  & 24   & 37   & 40   & 55   & 63  & 65   & 69   & 83  & 97   & 105  & 106  & 110  & 114  & 118  & 122  \\ \midrule
Number                 & 18k         & 41k    & 58.7k   & 145k  & 275k & 350k & 404k & 331k & 93k & 133k & 146k & 71k & 129k & 188k & 182k & 102k & 238k & 134k & 107k \\ \bottomrule
\end{tabular}%
}
\end{table}

The number of triangles (which is equivalent to the number of Gaussians) is set to be the same as Relit3DGS~\cite{relit3dgs} as listed in Table \ref{tab:ngaussian}. For MipNeRF-360 scenes, we employ 1M triangles (including foreground and background mesh) for all scenes as in SuGaR~\cite{sugar}.

\subsection{BRDF Function}
The BRDF function $f$ is defined in Eqn. \ref{eqn:brdf} consisting of a diffuse and a specular component.
\begin{equation} \label{eqn:brdf}
f(\omega_o, \omega_i) = \frac{1-m}{\pi}\bm{a} + \frac{DFG}{4(\omega_i \cdot \bm{n})(\omega_o \cdot \bm{n})}
\end{equation}

We use roughness remapping $\alpha=r^2$ and compute the half vector as $\bm{h}=(\bm{\omega}_i + \bm{\omega}_o) / \norm{\bm{\omega}_i + \bm{\omega}_o}$. The normal distribution function $D$ is defined as follows:
\begin{equation}
D(\alpha, \bm{h}, \bm{n}) = \frac{\alpha^2}{\pi((\bm{h}\cdot\bm{n})^2 (\alpha^2-1)+1)^2}
\end{equation}
The Fresnel reflection is defined as follows:
\begin{equation}
F(\bm{h}, \bm{n}) = F_0 + (1-F_0) (1-\bm{h}\cdot\bm{n})^5
\end{equation}
where $F_0 = 0.04(1-m) + \bm{a}m$.

We employ the correlated Smith masking-shadowing function defined as follows:
\begin{equation}
G(\alpha, \bm{n}, \bm{\omega}_i, \bm{\omega}_o) = \frac{1}{1 + G_{lmda}(\alpha, \bm{n}\cdot\bm{\omega}_i) + G_{lmda}(\alpha, \bm{n}\cdot\bm{\omega}_o)}
\end{equation}
where the lambda GGX masking function is defined as follows:
\begin{equation}
G_{lmda}(\alpha, \bm{n}\cdot\bm{\omega}_i)=\frac{1}{2(\sqrt{1+\alpha*(1-(\bm{n}\cdot\bm{\omega}_i)^2)/(\bm{n}\cdot\bm{\omega}_i)^2}-1)}
\end{equation}

\subsection{Hyperparameter Settings}
The strengths of base color, metallic, and roughness smoothness regularizer ($\lambda_{smooth}$ values) are set to 0.06, 0.02, and 0.02, respectively. We give much larger importance to the other regularizers ($\lambda_{o}, \lambda_{sc}$, and $\lambda_{sr}$ are all set to 1.0) since we observe that if a single Gaussian happens to behave abnormally, the whole training process becomes unstable in the following steps.
\begin{enumerate}
\item A vertex is driven to the opposite side of its original part of the surface, leading to very sharp and long triangles intersecting with other parts of the surface mesh. Or, a Gaussian is updated to have a large scale so that it is rendered outside the object region.
\item These erroneous Gaussians interfere especially with visibility. The global illumination shed to nearby Gaussians is mistakenly occluded.
\item The optimizer tries to compensate by either (1) increasing the intensity of the local illumination while dimming the global illumination completely, or (2) decreasing their opacity to the extreme.
\item The global illumination SH coefficients or some opacity values receive a large gradient and an overflow happens.
\end{enumerate}

\begin{table}[]
\centering
\caption{Learning rate values for each learnable parameter of our model.}
\label{tab:lr}
\resizebox{\textwidth}{!}{%
\begin{tabular}{@{}l|c|c|c|c|c|c|c|c|cc|cc|c@{}}
\toprule
\multirow{2}{*}{Parameter} & \multirow{2}{*}{Position} & \multirow{2}{*}{Color SH} & \multirow{2}{*}{Scale} & \multirow{2}{*}{Opacity} & \multirow{2}{*}{Base Color} & \multirow{2}{*}{Metallic} & \multirow{2}{*}{Roughness} & \multirow{2}{*}{Normal Rotation} & \multicolumn{2}{c|}{Direct Light SH} & \multicolumn{2}{c|}{Indirect Light SH} & \multirow{2}{*}{Visibility SH} \\ \cmidrule(lr){10-13}
                           &                           &                           &                        &                          &                             &                           &                            &                                  & deg=0       & deg\textgreater{}0     & deg=0       & deg\textgreater{}0       &                                \\ \midrule
LR                         & 1.6e-4                    & 2.5e-3                    & 5e-3                   & 5e-2                     & 1e-2                        & 1e-2                      & 1e-2                       & 1e-3                             & 2.5e-3      & 2.5e-4                 & 1e-3        & 1e-4                     & 0.1                            \\ \bottomrule
\end{tabular}%
}
\end{table}

The learning rate values for the Gaussian parameters and direct illumination SH coefficients are listed in Table \ref{tab:lr}. We use Adam~\cite{adam} optimizer with $(\beta_1, \beta_2)=(0.9, 0.999)$ and $\epsilon=1e-15$. 


\subsection{Relighting Evaluation}
It is well known that optimization of inverse rendering problems always comes with material-lighting ambiguity. Therefore, we explicitly compute 3-channel rescale factor for the base color by taking the median value of all pixel-wise ratios between our model's prediction and ground truth, following common practice~\cite{nerfactor,invrender,tensoir}. However, in the case of real captures, we omit this procedure and present only qualitative results as no ground truth on base color is provided.

The PBR rendering pipeline is in linear (HDR) space and the final result is clipped to $[0, 1]$ and mapped to SRGB just before calculating photometric loss or evaluation.

\section{Discussion on Potential Negative Impact}
We make clear that only publicly available datasets are employed for all our experiments.
In a case where personal photos are leaked, a person with malicious intentions can make use of our proposed model to efficiently produce a 3D model (such as face, body, or private space) from them.

\pagebreak

%
%
\bibliographystyle{splncs04}
\bibliography{main}